\documentclass[a4paper,11pt,openbib,final,footheight=-5pt]{scrartcl}
\usepackage[a4paper,includehead,top=20mm,bottom=20mm,right=20mm,left=20mm]{geometry}

\pdfoutput=1
\usepackage[breaklinks,pdftex]{hyperref}

\hypersetup{
pdftitle={Amirian, Jafarzadeh, Abali, Reali, Hogan},
pdfauthor={Benhour Amirian, Hossein Jafarzadeh, Bilen Emek Abali, Alessandro Reali, James David Hogan},
pdfkeywords={Phase-field model; Anisotropic brittle solids; Single crystals; Finite element method; Fracture and twinning; Monolithic scheme; FEniCS},
}

\makeatletter
\DeclareOldFontCommand{\bf}{\normalfont\bfseries}{\mathbf}
\DeclareOldFontCommand{\it}{\normalfont\itshape}{\mathit}
\makeatother

\usepackage[utf8]{inputenc} 

\usepackage[T1]{fontenc}
\usepackage[american]{babel}

\usepackage{url}            
\usepackage{booktabs}       
\usepackage{amsfonts}       
\usepackage{nicefrac}       
\usepackage{microtype}      
\usepackage{graphicx}
\usepackage{doi}
\usepackage{romannum}
\usepackage{xfrac}
\usepackage{units}
\usepackage{enumerate}
\usepackage{setspace}
\usepackage{chemformula}
\usepackage{amsmath, amsthm, amsxtra}
\usepackage{amssymb}
\usepackage{mathrsfs}
\usepackage{miller}

\usepackage{bm, upgreek}
\usepackage{booktabs, multirow}
\usepackage{subfigure}
\usepackage{float}
\usepackage{xcolor}

\newcommand{\bec}{\begin{center}}
\newcommand{\eec}{\end{center}}

\newcommand{\ten}[1]{\bm{#1}}
\newcommand{\p}{\partial}

\newcommand{\del}{\updelta}
\newcommand{\pd}[2]{\frac{\p #1}{\p #2}}
\newcommand{\dd}{\, \mathrm d }
\newcommand{\Dt}[1]{#1^{\scalebox{0.4}{\textbullet} } }

\newcommand{\begeq}{\begin{equation}\begin{gathered}}
\newcommand{\eqend}{\end{gathered}\end{equation}}
\newcommand{\begal}{\begin{equation}\begin{aligned}}
\newcommand{\alend}{\end{aligned}\end{equation}}

\usepackage{pdflscape}
\usepackage{fancyhdr} 
\fancypagestyle{mylandscape}{
\fancyhf{} 
\fancyfoot{
\makebox[\textwidth][r]{
  \rlap{\hspace{0.1cm}
    \smash{
      \raisebox{4.87in}{
        \rotatebox{90}{\thepage}}}}}}
}

\title{\huge Thermodynamically-consistent derivation and computation of twinning and fracture in brittle materials by means of phase-field approaches in the finite element method}

\author{Benhour Amirian$^{a}$\thanks{Corresponding author: benhour@ualberta.ca}
\and
Hossein Jafarzadeh$^{b}$
\and
Bilen Emek Abali$^{c}$\thanks{Corresponding author: bilenemek@abali.org}
\and
Alessandro Reali$^{b}$
\and
James David Hogan$^{a}$
\and
\\
\small $^a$Department of Mechanical Engineering \\[-0.1in]
\small University of Alberta, Edmonton, AB T6G 2R3, Canada
\\[0.02in]
\small $^b$Department of Civil Engineering and Architecture \\[-0.1in]
\small University of Pavia, I-27100 Pavia, Italy
\\[0.02in]
\small $^c$Department of Materials Science and Engineering, Division of Applied Mechanics\\[-0.1in]
\small Uppsala University, Ångströmlab Box 35, 751 03 Uppsala, Sweden
}

\date{} 

\begin{document}
\pagenumbering{arabic}

\maketitle

\begin{abstract}
A theoretical-computational framework is proposed for predicting the failure behavior of two anisotropic brittle materials, namely, single crystal magnesium and boron carbide. Constitutive equations are derived, in both small and large deformations, by using thermodynamics in order to establish a fully coupled and transient twin and crack system. To study the common deformation mechanisms (e.g., twinning and fracture), which can be caused by extreme mechanical loading, a monolithically-solved Ginzburg--Landau-based phase-field theory coupled with the mechanical equilibrium equation is implemented in a finite element simulation framework for the following problems: (i) twin evolution in two-dimensional single crystal magnesium and boron carbide under simple shear deformation; (ii) crack-induced twinning for magnesium under pure mode \Romannum{1} and mode \Romannum{2} loading; and (iii) study of fracture in homogeneous single crystal boron carbide under biaxial compressive loading. The results are verified by a steady-state phase-field approach and validated by available experimental data in the literature. The success of this computational method relies on using two distinct phase-field (order) parameters related to fracture and twinning. A finite element method-based code is developed within the Python-based open-source platform FEniCS. We make the code publicly available and the developed algorithm may be extended for the study of phase transformations under dynamic loading or thermally-activated mechanisms, where the competition between various deformation mechanisms is accounted for within the current comprehensive model approach.
\end{abstract}


\section{Introduction}

Understanding and predicting anisotropic fracture and damage evolution in brittle materials have been long-standing problems in engineering designs. Owing to the advent of novel modeling techniques and the advances in computational capabilities, the usage of accurate and robust numerical methods plays a key role in situations where purely experimental approaches are of high cost and not always readily accessible (e.g., high-energy in situ X-ray computed microtomography \cite{flannery1987three}, in situ electron backscattered diffraction (EBSD) \cite{wang2018situ}, and micro/nano-mechanical testing \cite{dehm2018overview}). In the literature, the simulation of fracture in solids at the atomic scale is commonly treated by molecular dynamics \cite{rountree2002atomistic}, density functional theory \cite{sharmainter}, or lattice static models that are based on spring networks \cite{thomson1992lattice}. Despite addressing nonlinearities at the crack tip, avoiding singularity-related issues, and considering bond breaking between atoms \cite{ferdous2017mode, patil2016comparative, chen2016nonlocal}, there exist challenges in using atomistic models to cover the time- and length-scales necessary to analyze the structural response at the macroscale needed for engineering applications.

Conventionally, there are two main categories of numerical approaches that are employed to provide explicit simulations of material failure: (i) discrete crack models (e.g., the discrete element method \cite{scholtes2012modelling, sinaie2018discrete}, the extended finite element method (XFEM) \cite{moes1999finite}, the cohesive zone method \cite{xu1994numerical}, and the cohesive segment methods \cite{remmers2003cohesive}) in which the displacement field is allowed to be discontinuous across the fracture surfaces, and (ii) smeared (continuum) crack models (e.g., damage models \cite{pijaudier1987nonlocal}, and diffuse interface models \cite{hakim2009laws,aldakheel2019virtual}) that consider a continuous displacement everywhere, assuming gradually decreasing stiffness to model the degradation process. Regardless of showing much success in modeling crack propagation \cite{linder2013strong}, the discrete crack models need additional criteria based on stress, strain energy density, energy release rate, or virtual crack closure techniques \cite{baydoun2012crack} to predict the crack initiation (nucleation), growth, and branching in dynamic fracture problems \cite{belytschko2003dynamic}. Further, the sharp representation of cracks requires remeshing algorithms or using the partition of unity method \cite{babuvska1997partition}, both having their own difficulties in tracking the multiple crack fronts in complex three-dimensional morphologies \cite{ingraffea1985numerical, sutula2018minimum}.

In the smeared crack approach, regularizing strong discontinuities caused by strain localizations within a finite and thin band leads to a precise approximation of the crack topology \cite{lorentz2003analysis}. The gradient damage model \cite{pham2010approche,aslan2011micromorphic}, physical/mechanical community-based phase-field fracture models \cite{karma2001phase, hofacker2012continuum, borden2012phase} that traces back to the reformulation of Griffith's principle \cite{griffith1921vi}, and peridynamics \cite{piola1846intorno, silling2000reformulation, dell2015origins}, which may be regarded as generalized non-local continuum mechanics, fall within this category. Replacing partial differential equations in the phase-field model by integrals in peridynamics allows for topologically-complex fractures such as intersecting and branching to be handled in both two and three dimensions \cite{roy2017peridynamics}. Coupling smeared and discrete crack approaches, for example, the element deletion method \cite{song2008comparative}, the combined non-local damage and cohesive zone method \cite{borst2004discrete,della2017modeling}, and the thick level-set method \cite{bernard2012damage} have also shown promising results in modeling fracture. In the thick level-set method, a discontinuous crack description is surrounded by continuous strain-softening regions, defined by a level-set function to separate the undamaged from the damaged zone \cite{moreau2015explicit}. However, the dependence of the results on the finite element mesh and the convergence path of the solutions, for a mesh size tending to zero, results in numerical errors \cite{mariani2003extended}.

\subsection{Phase-field approach} 

As an alternative approach, the phase-field model has been widely used recently in the context of phase transition processes, ranging from solidification \cite{boettinger2002phase} and phase transformation in solids \cite{schmitt2014crystal} to the modeling of ferroelectric materials \cite{schrade2007domain}. Having the capability to model the microstructural evolution, phase-field modeling has been successfully adopted in the simulation of martensitic phase transformations \cite{levitas2002three,javanbakht2021effect}, reconstructive phase transformations \cite{denoual2010phase}, phase transformations in liquids \cite{slutsker2006phase}, dislocations \cite{albrecht2020phase}, twinning \cite{clayton2011phase}, damage \cite{loew2019rate}, and their interactions \cite{ruffini2015phase, mozaffari2016coupled, schmitt2015combined}. Initiated with the celebrated work by Francfort and Marigo on the variational approach to brittle fracture \cite{francfort1998revisiting}, where the total energy is minimized simultaneously with respect to the crack geometry and the displacement field, the concept of applying the phase-field method in fracture mechanics has gained significant interest in the literature \cite{bourdin2000numerical, bourdin2008variational, farrahi2020phase, levitas2018thermodynamically, jafarzadeh2019phase, kuhn2010continuum, placidi2020variational, sargado2018high, eid2021multiscale, msekh2015abaqus, shanthraj2017elasto, aldakheel2018phase, selevs2021general}. Due to the thermodynamic driving forces, the evolution of interfaces (e.g., merging and branching of multiple cracks) is predicted with no additional effort \cite{wu2018phase}. Also, being quantitative, material-specific, and simple to couple to other calculations (e.g., stress or temperature \cite{cui2021phase}) makes phase-field modeling a powerful and flexible method for studying the fracture of single crystalline \cite{clayton2016phase} and polycrystalline materials \cite{emdadi2021phase} as well as in granular solids \cite{timofeev2021hemivariational}. The high computational cost in the phase-field model due to resolving the gradient term by using sufficiently refined mesh in the damaged zone is straightforwardly tackled by parallel implementations \cite{chen2019fft} and adaptive remeshing \cite{yue2006phase}.

Using the phase-field model to study the failure mechanisms in brittle materials has recently received increasing attention \cite{kasirajan2020phase, carrara2020framework, raghu2020thermodynamically, kiendl2016phase, wang2020phase}. Brittle solids may fail along grain boundaries or cracks propagate along the constituent phases in the case of geomaterials \cite{xie2012experimental, misra1997measured}. A fourth-order model for the phase-field approximation in brittle materials leads to a unified model to simulate the mechanics of damage and failure in concrete \cite{wu2017unified}, where an explicit Hilber--Hughes--Taylor-$\alpha$ \cite{hilber1977improved} method is used with a phase-field approach given by a single-well energy potential to describe the fracture behavior. This method is well-established in the mechanics community. In the physics community, on the other hand, the phase-field models are commonly derived by adapting the phase transition formalism of Landau and Ginzburg \cite{landau1980statistical}. For example, Aranson et al. \cite{aranson2000continuum} combined elastic equilibrium with the Ginzburg--Landau equation, which accounted for the dynamics of defects, to study the crack propagation in brittle amorphous solids. Another Ginzburg--Landau based phase-field approach restricted to mode \Romannum{3} fracture (antiplane shear) was proposed by Karma et al. \cite{karma2001phase} and Hakim and Karma \cite{hakim2009laws} in the two- and three-dimensional settings, respectively. Considering fracture as a solid-gas transformation, the double-well energy potential appeared in phase-field modeling of damage \cite{levitas2018thermodynamically}. Some of the disadvantages of the double-well potential, such as crack widening and lateral growth during crack propagation, can be eliminated by using a single-well term \cite{levitas2021review}; however, the realistic shape of the stress-strain curves obtained from the experiments or atomistic simulations are more difficult to capture by the single-well free energy density \cite{farrahi2020phase}.

In numerical implementations, nonlinear problems with a strong coupling between the equilibrium equation and the phase-field parameter is solved through two approaches: 
\begin{enumerate}
\item
A staggered solution scheme is based on decoupling balance equations and the phase-field problem into the system of two equations that are solved in a subsequent manner \cite{miehe2010thermodynamically, miehe2010phase}. The implementation is more modular and two smaller systems to solve is faster than two systems together; computational time increases exponentially. The method is robust due to giving rise to two convex minimization problems, but depending on the coupling (and application), a significant amount of staggered iterations may be required at a fixed loading step, thus resulting in a higher computational cost \cite{singh2016fracture}.
\item
A monolithic solution where all variables are solved at once (simultaneously) \cite{heister2015primal}. In some cases, for example highly coupled systems, the monolithic solution is more efficient as a result of (in total) less Newton--Raphson iterations \cite{ambati2015review}. To the best of our knowledge, no studies have focused on solving the Ginzburg--Landau based phase-field problem concerned with predicting the twinning and fracture behavior of brittle materials by using a monolithic scheme; this is addressed in the current paper.
\end{enumerate}

\subsection{Goals and outlook} 

In the present study, we seek to extend the Ginzburg--Landau phase-field approach to predict fracture and twinning in single crystal anisotropic brittle materials (e.g., magnesium, \ch{Mg}, and boron carbide, \ch{B4C}). The advantage of the Ginzburg--Landau approach over the incremental energy minimization method used in Clayton and Knap \cite{clayton2016phase} is that material parameters associated with time scales for interfacial motion enter the model by calibrating it with the most recent Molecular Dynamics simulations \cite{hu2020disconnection}, making it a more general model for studying the deformation mechanisms of intrinsically brittle materials (e.g., single crystalline \ch{Mg} and \ch{B4C}). This is an important consideration for deformation twinning since the propagation speed of twin boundaries can be difficult to measure, and could even be supersonic if the driving stress is sufficiently large \cite{rosakis1995dynamic}. In addition, this work focuses on derivation of governing equations and solving them monolithically in order to increase the accuracy for applications with strong coupling between mechanics and damage. We solve the nonlinear and coupled differential equations by using the open-source parallel computing platform FEniCS \cite{alnaes2015fenics}. By following the works on local stress concentrations in nanoscale defect-free volumes or by high pressures \cite{levitas2005thermomechanical}, as well as shear arising from twinning \cite{clayton2010modeling}, we develop a nonlinear phase-field theory for elasticity along with anisotropic surface energy \cite{gorbushin2020stress}. To address this, the governing equations for both small and large deformations are derived because finite rotations may occur under some loading conditions even at small strains, which necessitate considering both regimes in crystallographic theory \cite{roters2010overview}. A new decomposition for the strain energy density based on \cite{van2020strain} is proposed to reproduce the experimentally-observed crack propagation under compressive loading, and the simulated results are compared with analytical solutions. In these comparisons, a double-well energy potential is considered for studying the fracture behavior (e.g., crack initiation, growth, and propagation) and twinning in anisotropic  brittle solids.

The remainder of this paper is outlined as follows. In Section \ref{section.materials}, we briefly describe the chosen materials. Continuum mechanics and thermodynamically sound derivation of equations are shown in Section \ref{section.formulation}. Variational formulation and the finite element method is explained in Section \ref{section.computation}. Results and representative material properties along with the discussion of phase-field simulations are reported in Section \ref{section.results}. The conclusions of the study are drawn in Section \ref{section.conclusion}. 

\section{Materials}\label{section.materials}

The focus of the present paper is to model the deformation behavior of magnesium and boron carbide single crystals. The low ductility of these materials suggests to comprehend them as brittle elastic materials with large driving forces for dislocation glide, leading to other mechanisms such as phase transformations \cite{edalati2011plastic}, deformation twinning \cite{arlt1990twinning,forest2001strain}, and fracture \cite{conrad2000electroplasticity}.

\subsection{Magnesium}

Having low mass density ($\sim$23\% of steel and 66\% of aluminum), high strength, and durability for a wide range of temperatures in high performance automotive and aerospace applications, magnesium and its alloys have attracted considerable attention in recent years \cite{kainer2000magnesium, mordike2001magnesium}. \ch{Mg} alloys tend to be brittle due to their limited number of dislocation systems \cite{rohrer2001structure}. As a result of possessing low-symmetry crystallographic structure and larger critical resolved shear stress (CRSS), twinning is the dominant deformation mode in magnesium as it is subjected to twinning-favored loads stretched along [0001] \cite{beyerlein2010statistical, brown2012role, morrow2016characterization}, resulting in transitions in the material behavior at high strain rates \cite{gong2015prismatic}. A previous study indicated that the formation of intersecting twins may improve the ductility of \ch{Mg} alloys \cite{lentz2016strength}. Therefore, understanding and predicting the twinning behavior during plastic deformation of magnesium is critical towards the realization of next-generation lightweight metallic materials for application in automotive and defense applications. In order to investigate twinning in magnesium, various techniques such as high-resolution transmission electron microscopy \cite{sim2018anomalous}, visco-plastic self-consistent polycrystal models \cite{lebensohn1993self}, elasto-plastic self-consistent polycrystal models \cite{turner1994study}, molecular dynamics simulations \cite{hu2020embracing}, crystal plasticity models \cite{lindroos2017effect}, and quasi-static phase-field models \cite{clayton2011phase} have been employed. In this paper, the fracture and twinning behaviors of single crystal magnesium are studied using an advanced time-dependent phase-field theory by numerically solving engineering problems.

\subsection{Boron carbide}

As a result of possessing hardness above $\unit[30]{GPa}$, low mass density ($\unit[2.52]{\sfrac{g}{cm^{3}}}$), and high Hugoniot elastic limit ($\unit[17]{}$-$\unit[20]{GPa}$), boron carbide (\ch{B4C}) has received considerable attention in ballistic applications \cite{karandikar2009review}. Due to its high melting point and thermal stability \cite{matkovich1977boron}, favorable abrasion resistance \cite{subramanian2010development}, and high temperature semiconductivity \cite{thevenot1990boron}, boron carbide excels in refractory, nuclear, and novel electronic applications, respectively; however, its performance is hindered by one or more of a number of inelastic deformation mechanisms, including deformation twinning \cite{li2010deformation}, stress-induced phase transformations \cite{an2014atomistic,eremeyev2010phase}, and various fracture behaviors \cite{an2015atomistic} when subjected to mechanical stresses exceeding their elastic limit. The key failure mechanisms in boron carbide (e.g., cleavage fracture and twinning) are commonly studied experimentally using numerous characterization techniques (e.g., transmission electron microscopy \cite{zhao2016directional} and Raman spectroscopy \cite{yan2009depressurization}). Fracture in the form of shear failure, cavitation, and cleavage has been confirmed from atomic simulation results, either via first principles or molecular dynamics simulations \cite{fanchini2006behavior, taylor2012effects}. Finite deformation continuum models, such as cohesive zone models for fracture \cite{clayton2005dynamic} and crystal plasticity \cite{padilla2017coupled} have also been used to investigate inelastic deformation in single and polycrystalline boron carbide. The present time-evolved phase-field model seeks to engineer the next generation of anisotropic boron carbide-based armor ceramics by understanding the important plastic deformation and brittle fracture mechanisms that govern its high rate performance. The current framework does not incorporate slip for \ch{B4C} due to having very large resistance to dislocation glide in certain directions at low temperatures \cite{hirth1983theory}.

\section{Formulation} \label{section.formulation}

In this section, we develop a model for single twinning and fracture systems in solids based on thermodynamical derivation. The present approach extends the model of Clayton and Knap \cite{clayton2011phase, clayton2016phase, clayton2013phase, clayton2015phase} by accounting for the time-evolution of order parameters towards an equilibrium state for predicting the twinning and crack paths in anisotropic single crystal materials adequately. This allows the study of spatio-temporal fluctuations of order parameters (e.g., twinning and fracture variables), as well as the nanoscale dynamics that govern various pattern forming phenomena \cite{provatas2011phase}. Moreover, the interfaces, their propagation, and interactions, which are the most important features governing the formation of microstructures in materials, can be studied via this newly implemented approach. It is worth mentioning that the present work does not address plastic slip distinct from the motion of twinning partials inherent in deformation twinning. We refer to \cite{clayton2005geometric, clayton2010nonlinear} and references therein for further details on ways to simultaneously address plastic slip and twinning. In what follows, we are interested in elastic twinning (i.e., the reversible nature of the corresponding deformation of the crystal) in which a twin appears and grows in the crystal lattice due to the presence and to the increase of an external load, and escapes from the crystal upon removal of the load \cite{klassen2012mechanical}.

\subsection{Order parameters}

The main desired feature of the proposed model is to introduce an order parameter $\eta$ assigned to each material point $\ten{X}$ for the description of the twinning, $\eta= 0$ within the parent (original or untwinned phase at $(\ten{X}, t)$) elastic crystal, whereas $\eta=1$ denotes the twin. The twin boundary zone is determined by the diffuse interval, $0<\eta<1$. Another order parameter, $\xi$, is used to represent fracture, where $\xi=0$ indicates undamaged (``virgin'') material, $\xi=1$ fully damaged material, and $\xi \in (0,1)$ partially degraded material. Both of these state variables are commonly assumed to be at least $\mathcal{C}^{2}$-continuous with respect to $\ten{X}$ according to the diffuse interface theory \cite{cahn1958free, allen1979microscopic}. They also vary in time and are subjected, in general, to time-dependent boundary conditions. In addition, the global irreversibility constraint of crack evolution is satisfied by ensuring locally a positive variational derivative of the crack surface function and a positive evolution of the crack phase field \cite{aranson2000continuum}. Without this fundamental constraint, no cyclic loading can be performed \cite{placidi2020variational, timofeev2021hemivariational}; however, cyclic loading is not in the scope of the current study.

\subsection{Kinematics}

\begin{figure}
\centering
\includegraphics[width=100mm, height=75mm]{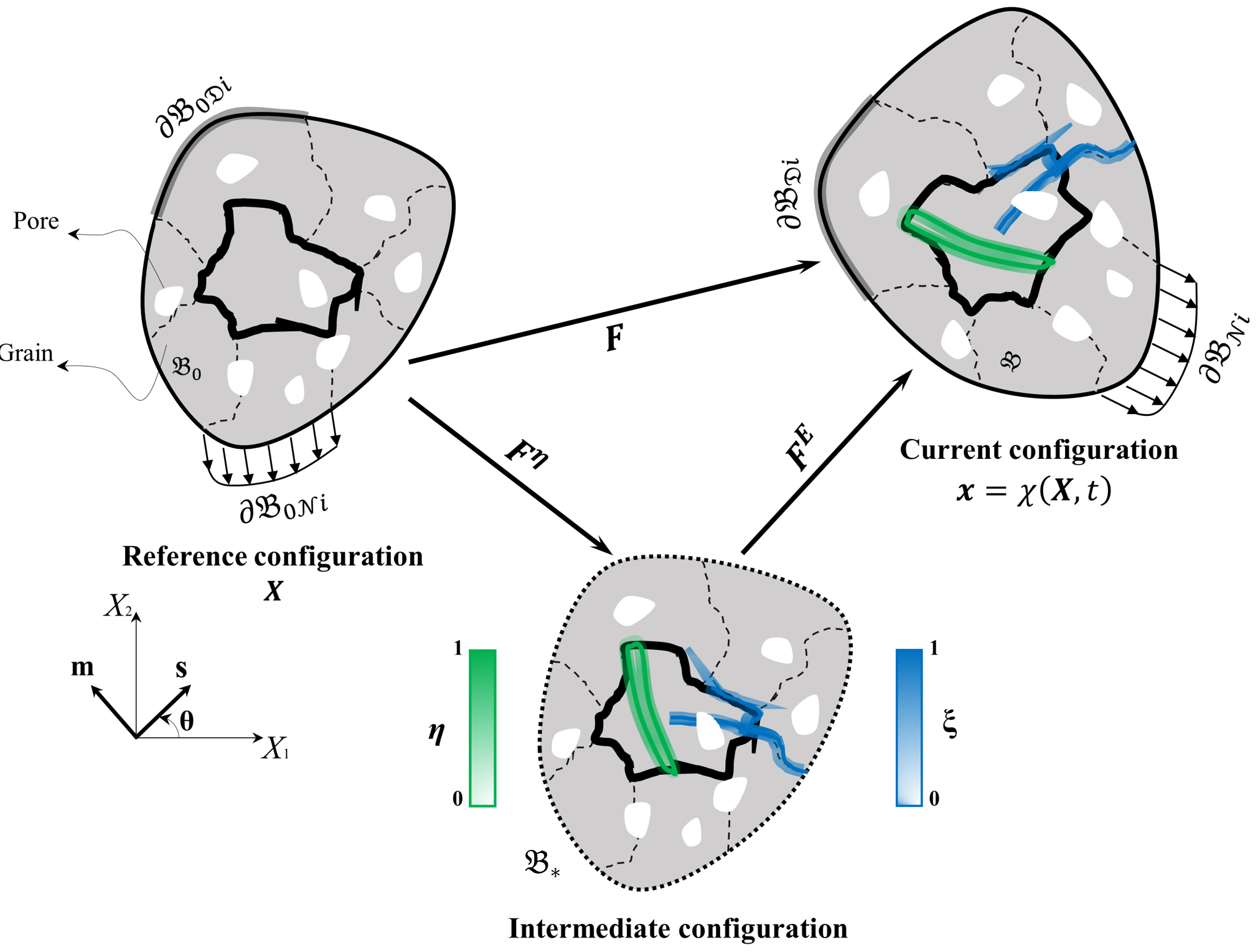}
\caption{Multiplicative decomposition of the deformation gradient into elastic $\boldsymbol{F}^{\text{E}}$ and stress-free twinning shear $\boldsymbol{F}^{\eta}$ parts with corresponding configurations. The reference configuration ${\mathfrak{B}_{0}}$, the deformed configuration ${\mathfrak{B}}$, and an arbitrary elastically unloaded intermediate configuration ${\mathfrak{B}_{*}}$ of a polycrystal material are shown. The Neumann and Dirichlet boundary conditions corresponding to each configuration are also illustrated. Two distinct order parameters for fracture $\xi$ (blue color) and twinning $\eta$ (green color) are also considered. (For interpretation of the references to color in this figure, the reader is referred to the online version of this article.)}
\label{fig1}
\end{figure}

We use standard continuum mechanics notation and understand a summation over repeated indices. A continuum body as in Figure \ref{fig1}, composed of many grains, is considered in the reference frame, $\mathfrak{B}_{0} \subset \mathbb{R}^{d}$ in $d$ dimensional space. This framework is simply the initial configuration---before loading. For modeling purposes, we introduce a stress-relaxed intermediate frame, ${\mathfrak{B}_{*}}$, and a current frame, ${\mathfrak{B}}$. Since the formulation is established in the reference frame, it is a material system where coordinates denote particles. In this frame, the mass density is given as a function in $\ten X$, we circumvent solving the balance of mass. For the balance of momentum, the computational domain will be ${\mathfrak{B}_{0}}$ with its closure $\partial{\mathfrak{B}_{0}}$. On Neumann boundaries, $\partial{\mathfrak{B}_{0_{\mathcal{N}i}}}$, gradient of the solution (traction vector) is known and on Dirichlet boundaries, $\partial{\mathfrak{B}_{0_{\mathcal{D}i}}}$, the solution (displacement) is given. 
The motion from the reference position $\ten X$ to the current (deformed) position $\ten x = \ten X + \ten u$ is given by the displacement tensor of rank one, $\ten u=\ten u(\ten X, t)$, as a function in space, $\ten X$, and time, $t$. The deformation gradient, $\ten F = \nabla_{0}\ten{x}$, is multiplicatively decomposed as
\begin{eqnarray}\label{eq.1}
F_{ij} = \pd{x_i}{X_j} = x_{i,j} = F^{\text{E}}_{ik} F^{\eta}_{kj},
\end{eqnarray} 
where $\ten F^{\text{E}}$ is the recoverable elastic deformation work-conjugate to Piola stress, analogously, $\ten F^{\eta}$ is the deformation associated with structural defects, twinning in the current study, evolving within the material. It is important to highlight that twinning is distinguished from plastic slip. First, the former occurs by collective motion of defects that preserves the particular orientation relationship between the twin and original phases \cite{christian1995deformation}. In addition, twinning is unidirectional while usually slip is not \cite{zambaldi2015orientation}. In contrast to $\ten{F}$, which always satisfies compatibility conditions $\nabla_{0} \times \ten{F}=0$, the deformation maps $\ten{F}^{\text{E}}$ and $\ten{F}^{\eta}$ are generally not integrable to a vector field, when considered individually, as a result of existing crystal defects \cite{bilby1955continuous, clayton2004anholonomic}. In other words, these two deformations are generally anholonomic ($\nabla_{0} \times \ten{F}^{\eta} \neq 0$ and $\nabla \times {(\ten F^\text{E})}^{-1} \neq 0$) \cite{clayton2012anholonomic, clayton2014differential}. Additional structural changes may be included in this theory for representation of other defects, such as point defects \cite{clayton2009non} or dislocation slips \cite{kondo2014phase,forest2000material}. The kinematics for twinning in simple shear is given as \cite{james1981finite}
\begal\label{strain.def}
\ten{F}^{\eta}(\eta) = \ten{I} + \phi(\eta)\gamma_{0} \ten{s} \otimes \ten{m} \ ; & \  F_{ij}^{\eta}=\delta_{ij}+\phi(\eta)\gamma_{0} s_{i} m_{j},
\alend 
where $\ten{s}$ and $\ten{m}$ are the orthogonal unit vectors in the directions of twinning and normal to the twinning plane, respectively; and $\gamma_{0}$ is the magnitude of the maximum twinning shear. All functions are defined in the reference frame, since we have a material system. For small deformation, this distinction is negligible. The continuous interpolation function $\phi(\eta)$ is obtained from a general representative function $\varphi(a,\eta)$ within a fourth-degree potential \cite{levitas2002three} defined as
\begin{eqnarray}
\varphi(a,\eta)=a\eta^{2}(1-\eta)^2+\eta^{3}(4-3\eta),
\end{eqnarray}
where $a$ is a constant parameter---in order to ensure that $\varphi(a,\eta)$ is a monotonous function, $a$ is chosen between 0 and 6. The functions $\varphi(a,\eta)$ and $\phi(\eta)$ should be monotone for $0 \leq \eta \leq 1$ and satisfy the following conditions \cite{levitas2002three}
\begeq\label{interpolation.varphi}
\varphi(a,0)=0, \ \varphi(a,1)=1, \ \pd{\varphi(a,0)}{\eta}=\pd{\varphi(a,1)}{\eta}=0, \\
\phi(\eta = 0) = 0, \ \phi(\eta = 1)=1, \ \pd{\phi(\eta = 0)}{\eta}=\pd{\phi(\eta = 1)}{\eta}=0.
\eqend
Setting $a = 3$ leads to $\phi(\eta) = \varphi(3, \eta) = \eta^{2}(3-2\eta)$, which obeys the antisymmetry condition, i.e., $\phi(1-\eta)=1-\phi(\eta)$ \cite{basak2018nanoscale}. As usually assumed, the plastic deformation is deviatoric such that the volume remains the same, $\det \ten{F}^{\eta}=1$. Therefore, the Jacobian determinant reads, $J=\det \ten{F}=\det \ten{F}^{\text{E}}$. With the right Cauchy--Green deformation tensors, $C_{ij}=F_{ki} F_{kj}$ and $C_{ij}^{\text{E}}=F_{ki}^{\text{E}} F_{kj}^{\text{E}}$, we obtain the Green--Lagrange total and elastic strains,
\begeq\label{def.strains}
E_{ij}=\frac{1}{2}(C_{ij}-\delta_{ij}),\ 
E_{ij}^{\text{E}}=\frac{1}{2}(C_{ij}^{E}-\delta_{ij}), 
\eqend
respectively. They are now used, as follows:
\begal\label{def.nonlin.strain}
E_{ij} = &
\frac12 \Big( C_{ij}-\delta_{ij} \Big)
= \frac12 \Big( F_{ki} F_{kj} - \delta_{ij} \Big) 
\\
= & \frac12 \Big( F^\text{E}_{kn} F^\eta_{ni} F^\text{E}_{km} F^\eta_{mj} - \delta_{ij} \Big)
= \frac12 \Big( C^\text{E}_{nm} F^\eta_{ni} F^\eta_{mj} - \delta_{ij} \Big) 
\\
= &\frac12 \bigg( C^\text{E}_{nm} \Big( \delta_{ni}+\phi(\eta)\gamma_{0} s_n m_i \Big) \Big( \delta_{mj}+\phi(\eta)\gamma_{0} s_m m_j \Big) - \delta_{ij} \bigg)
\\
= &\frac12 \bigg( 2 E^\text{E}_{ij} + C^\text{E}_{im} \phi(\eta)\gamma_{0} s_m m_j + C^\text{E}_{nj} \phi(\eta)\gamma_{0} s_n m_i + C^\text{E}_{nm} \phi^2 \gamma^2_{0} s_n s_m m_i m_j \bigg) 
\alend
In the case of small deformations, by using the standard linearization approach, we obtain
\begal \label{def.small.strain.approx}
E_{ij} = & 
\frac12 \Big( F_{ki} F_{kj} - \delta_{ij} \Big) 
= \frac12 \Big( ( \delta_{ki} + u_{k,i} ) ( \delta_{kj} + u_{k,j} ) - \delta_{ij} \Big) 
= \frac12 \Big( u_{j,i} + u_{i,j} + u_{k,i}u_{k,j} \Big)
\\
\approx & \varepsilon_{ij} =  \frac12 \Big( u_{j,i} + u_{i,j} \Big) \ .
\alend
With the same approach, by discarding all nonlinear terms (including displacement gradient multiplied by $\ten s$ or $\ten m$), we determine from Eq.\,\eqref{def.nonlin.strain}, in the case of small deformations,
\begeq
\varepsilon_{ij} =  \varepsilon^\text{E}_{ij} + \underbrace{ \frac12 \Big( \phi(\eta)\gamma_{0} s_i m_j + \phi(\eta)\gamma_{0} s_j m_i \Big) }_{\varepsilon^\eta_{ij} } \ .
\eqend
By inserting Eq.\,\eqref{def.small.strain.approx}, we obtain
\begeq 
\varepsilon^{\text{E}}_{ij}= 
\varepsilon_{ij} - \varepsilon^\eta_{ij} =
\frac{1}{2}\Big( u_{i,j} + u_{j,i} -\phi(\eta)\gamma_{0}\big( s_{i} m_{j} + m_{i} s_{j}\big) \Big) .
\eqend
In this way, we aim for using $\ten F$ and $\ten F^\eta$ in large deformation simulations; analogously, $\ten\varepsilon$ and $\ten\varepsilon^\eta$ in small deformation simulations. A comparison will lead to the significance of nonlinearity in twinning simulations.

\subsection{Constitutive equations}

In order to derive the constitutive equations, we follow thermodynamics of irreversible phenomena and refer to \cite{muller2009fundamentals} for historical remarks. By starting with the balance of energy and subtracting the balance of momentum, for an arbitrary volume $V_{0}$ in the undeformed configuration $\mathfrak{B}_{0}$, we obtain the balance of internal energy:
\begeq\label{bal.energy}
\int_{V_0} \rho_0 \Dt u \dd V + \int_{\p V_0} Q_i N_i \dd A - \int_{V_0} \rho_0 r \dd V = \int_{V_0} P_{ki} \Dt F_{ik} \dd V ,
\eqend
with the specific internal energy, $u$ in J/kg, its (heat) flux term, $\ten Q$ in W/m$^2$, across the direction $\ten N$ (surface normal outward the volume), a specific volumetric heat supply rate, $r$ in W/kg, and a production term defined by Piola stress, $\ten P$ in N/m$^2$, and deformation gradient rate, among others see \cite[Sect. 2.4]{027} for a straight-forward derivation. Temperature $T$ in K, and specific entropy $s$ in J/(K kg), are related by the global entropy balance equation, in the undeformed configuration, with the assumption that the entropy flux is $1/T$ times heat flux leading to 
\begeq
\int_{V_0} \rho_0 \Dt s \dd V + \int_{\p V_0} \frac1T Q_i N_i \dd A - \int_{V_0} \frac 1T \rho_0 r \dd V = \int_{V_0} \Sigma \dd V ,
\eqend
where the entropy production, $\Sigma$, is zero for reversible and positive for irreversible processes. This assertion, $\Sigma \geq 0$, is the Second Law of Thermodynamics. Since $T\geq 0$, we immediately acquire $T\Sigma\geq 0$. By replacing the supply term in Eq.\,\eqref{bal.energy} and using Gau\ss{}--Ostrogradskiy (divergence) theorem, we obtain
\begeq
\int_{V_0} \bigg( \rho_0 \Dt u + Q_{i,i} - \Big( \rho_0 T \Dt s + T \Big(\frac{Q_i}{T}\Big)_{,i} - T \Sigma \Big) \bigg) \dd V = \int_{V_0} P_{ki} \Dt F_{ik} \dd V  .
\eqend
Now, by using the Helmholtz free energy, $\psi = U - T s$, and using temperature as an independent thermodynamic parameter instead of entropy, the entropy production related term reads
\begal
\int_{V_0} T \Sigma  \dd V 
=&
\int_{V_0} \bigg( - \rho_0 \big( \Dt u - T \Dt s \big) - Q_{i,i} + T \Big(\frac{Q_i}{T}\Big)_{,i} + P_{ki} \Dt F_{ik} \Big) \bigg) \dd V 
\\
=&
\int_{V_0} \bigg( - \rho_0 \big( \Dt \psi + \Dt T s \big) + T Q_i \Big(\frac1{T}\Big)_{,i} + P_{ki} \Dt F_{ik}  \bigg) \dd V 
\\
=&
\int_{V_0} \bigg( - \rho_0 \big( \Dt \psi + \Dt T s \big) - \frac{Q_i}{T} T_{,i} + P_{ki} \Dt F_{ik}  \bigg) \dd V \ .
\alend
As the latter is positive, several restrictions are possible for constitutive relations. We choose the simplest possible constitutive relation for the heat flux. The linear dependency is called Fourier conduction law, $Q_i=-\kappa_{ij} T_{,j}$, where $\ten \kappa$ is the (symmetric and positive-definite) thermal conductivity tensor, it may depend on temperature \cite{clayton2010nonlinear} but is constant in temperature gradient. Hence, we reduce $T\Sigma \geq 0$ to the following inequality
\begeq\label{second.law2}
\int_{V_0} \bigg( - \rho_0 \big( \Dt \psi + \Dt T s \big) + P_{ki} \Dt F_{ik}  \bigg) \dd V \geq 0 \ .
\eqend
We start modeling by assuming the free energy dependency on order parameters and their first derivatives, $\psi=\psi(\ten F, T, \eta, \eta_{,i}, \xi, \xi_{,i})$, leading to
\begeq
\Dt \psi = \pd{\psi}{F_{ij}} \Dt{F_{ij}} 
+ \pd{\psi}{T}\Dt T 
+ \pd{\psi}{\eta} \Dt\eta
+ \pd{\psi}{\eta_{,i}} \Dt\eta_{,i}
+ \pd{\psi}{\xi} \Dt\xi
+ \pd{\psi}{\xi_{,i}} \Dt\xi_{,i} \ .
\eqend
By using the latter in Eq.\,\eqref{second.law2}, regrouping by means of variables in the energy $\ten F$, $T$, $\eta$, $\eta_{,i}$, $\xi$, and $\xi_{,i}$, we obtain
\begeq
\int_{V_0} \bigg( 
- \rho_0 \Big( s + \pd{\psi}{T} \Big) \Dt T
+ \Big( P_{kj} - \rho_0 \pd{\psi}{F_{jk}} \Big)  \Dt F_{jk} 
- \rho_0 \pd{\psi}{\eta} \Dt\eta
- \rho_0 \pd{\psi}{\eta_{,i}} \Dt\eta_{,i}
\\
- \rho_0 \pd{\psi}{\xi} \Dt\xi
- \rho_0 \pd{\psi}{\xi_{,i}} \Dt\xi_{,i}
\bigg) 
\dd V \geq 0 \ .
\eqend
This model is the simplest one and we obtain after subsequent Gau\ss{}--Ostrogradskiy theorems
\begal\label{second.law3}
\int_{V_0} \Bigg( &
- \rho_0 \Big( s + \pd{\psi}{T} \Big) \Dt T
+ \Big( P_{kj} - \rho_0 \pd{\psi}{F_{jk}} \Big)  \Dt F_{jk} 
\\
&+ \bigg(  - \rho_0 \pd{\psi}{\eta} + \rho_0 \Big( \pd{\psi}{\eta_{,i}}\Big)_{,i} \bigg) \Dt\eta
\\
&+ \bigg(  - \rho_0 \pd{\psi}{\xi} + \rho_0 \Big( \pd{\psi}{\xi_{,i}}\Big)_{,i} \bigg) \Dt\xi
\Bigg) 
\dd V
\\
- \int_{\p V_0} \Big(& \rho_0 \pd{\psi}{\eta_{,i}} \Dt\eta + \rho_0 \pd{\psi}{\xi_{,i}} \Dt\xi \Big) N_i \dd A  \geq 0 \ .
\alend
The boundary conditions for the evolution of the order parameters are obtained by assuming a phase-independent energy of the external surface $\partial V_{0}$ of $V_{0}$
\begeq\label{boundary}
\rho_0 N_i \pd{\psi}{\eta_{,i}} = 0,\
\rho_0 N_i \pd{\psi}{\xi_{,i}} = 0.
\eqend
With this assumption, fractured or twinned regions are always orthogonal to the boundary because their conjugate is proportional to the normal gradient. We refer to \cite{placidi2018strain} for different interpretations about the boundary conditions in the damage gradient approach. However, in the phase-field theory, one can use a stricter approach by introducing a generalized surface force conjugated to the rate of change of the order parameter. This generalized surface force balances the terms that appears due to the dependence of the free energy on the gradient of the order parameter, and this will give us more general boundary conditions \cite{levitas2013phase}. 
The inequality \eqref{second.law3} has to hold for any process, the first two terms may only be zero, since $\Dt T$ and $\Dt{\ten F}$ are in general not restricted and calculated by balance of entropy and momentum, respectively. Hence, we acquire the well-known relations
\begeq
s = - \pd{\psi}{T} , \ 
P_{kj} = \rho_0 \pd{\psi}{F_{jk}} \ ,
\eqend 
where the second term is known as Castigliano's theorem. In the case of small deformations, the latter differentiation reads $\sigma_{ij} = \rho_0 \p\psi/\p\varepsilon_{ji}$. The Second Law of Thermodynamics is satisfied for all processes by choosing  (mobility) parameters $\mathcal{L}^{\eta}$ and $\mathcal{L}^{\xi}$ in order to achieve Ginzburg--Landau (evolution) equations
\begal
\Dt\eta =& \mathcal L^\eta \Bigg( - \rho_0 \pd{\psi}{\eta} + \rho_0 \Big( \pd{\psi}{\eta_{,i}}\Big)_{,i} \Bigg) \ ,\\ 
\Dt\xi =& \mathcal L^\xi \Bigg( - \rho_0 \pd{\psi}{\xi} + \rho_0 \Big( \pd{\psi}{\xi_{,i}}\Big)_{,i} \Bigg) \ ,
\alend
which reduce Eq.\,\eqref{second.law3} to
\begeq
\int_{V_0} \bigg( \frac{ ( \Dt\eta )^2 }{\mathcal L^\eta} + \frac{ (\Dt\xi)^2 }{\mathcal L^\xi} \bigg) \dd V \geq 0 \ .
\eqend
The $\mathcal{L}^{\eta}$ and $\mathcal{L}^{\xi}$ are positive kinetic coefficients for twinning and fracture evolution, respectively.

\subsection{Governing equations}

For a given temperature, i.e. for an isothermal process, the balance of entropy is fulfilled and we aim for solving the balance of momentum and evolution equations for order parameters. We make further assumptions and model the system without inertia and body forces. In other words, for an isothermal, quasi-static case, the deformation is caused by the mechanical loading on boundaries such that the governing equations become
\begal \label{gov.eq1}
P_{ji,j} =& 0 \, , \ P_{ji} = \rho_0 \pd{\psi}{F_{ij}} \ , \\ 
 \frac{1}{\mathcal L^\eta} \Dt\eta =&  - \rho_0 \pd{\psi}{\eta} + \rho_0 \Big( \pd{\psi}{\eta_{,i}}\Big)_{,i}  \ ,\\ 
 \frac{1}{\mathcal L^\xi}  \Dt\xi =&  - \rho_0 \pd{\psi}{\xi} + \rho_0 \Big( \pd{\psi}{\xi_{,i}}\Big)_{,i}  \ .
\alend
As usual in mechanics, we search for displacement, $\ten u$, from the balance of momentum in Eq.\,\eqref{gov.eq1}$_1$. Equation \eqref{gov.eq1}$_2$ is solved in order to calculate $\eta$ for the twin versus original phase, and Eq.\,\eqref{gov.eq1}$_3$ is solved for determining $\xi$ for fracture versus intact material. Specific Helmholtz free energy, $\psi$, is simpler to model, if separated into mechanical and gradient of order parameters-related terms
\begeq
\psi(\ten F, \eta, \xi, \nabla \eta, \nabla \xi)
= g(\xi) \psi^\text{M}(\ten F, \eta) +
\psi^\nabla(\eta, \eta, \xi, \nabla \eta, \nabla \xi) \ .
\eqend
Technically, $\psi^{\nabla}$ represents the gradient energy per mass. The mechanical energy degrades by the order parameter $\xi$ denoting the microporosity ($\xi=1$ means fracture) of the structure in each position. This field function, $\xi$, is used to obtain a degradation function phenomenologically
\begeq\label{degradation}
g(\xi)=\zeta + (1-\zeta) \left(1-\xi \right)^2 ,
\eqend
The constant $\zeta$ ensures a minimal residual stiffness for fully fractured materials. The quadratic degradation of elastic energy has likewise been used in a number of other phase-field and gradient damage models \cite{bleyer2018phase, pham2011gradient, farrell2017linear, dammass2021unified}. The reflection or rotation of the reference frame of the crystal lattice commensurate with twinning should be taken into account for anisotropic elastic constants \cite{clayton2009continuum}. By using Green--Lagrange strains in Eq.\,\eqref{def.strains}, for the deformation energy density, $W^\text{M}$, we use a quadratic energy description 
\begeq
W^\text{M} = \rho_0 \psi^\text{M} = \frac{1}{2} E^\text{E}_{ij} C_{ijkl} E^\text{E}_{kl} .
\eqend
The stiffness tensor, $\ten C$, has minor and major symmetries, $C_{ijkl}=C_{jikl}=C_{klij}$. Also it depends on the twin stiffness $\ten C^\text{T}$ and initial stiffness $\ten C^\text{P}$ by the phase-field approach,
\begeq
\ten C = \ten C^\text{P} + ( \ten C^\text{T} - \ten C^\text{P} ) \phi(\eta) \ .
\eqend
The order parameter, $\eta$, is used to determine the amount of each phase in various position. Elastic coefficients of the fully twinned crystals, $\eta=1$, are related to those of the untwinned state, $\eta=0$, by
\begin{eqnarray}
C_{ijkl}^\text{T} = \mathcal{Q}_{im}\mathcal{Q}_{jn}\mathcal{Q}_{ko}\mathcal{Q}_{lp} C_{mnop}^\text{P} ,
\end{eqnarray}
where $\ten{\mathcal{Q}}$ is the reorientation matrix transforming the original lattice to twin lattice within a centrosymmetric structure
\begeq
\mathcal{Q}_{ij} =
    \begin{cases}
      2\textit{m}_{i} \textit{m}_{j} - \delta_{ij} & \text{type I twins},\\
      2\textit{s}_{i} \textit{s}_{j} - \delta_{ij} & \text{type II twins}.
    \end{cases}
\eqend
Type I and type II twins differ in reflections or rotations of the lattice vectors in the twin and parent phase. In the case of homogeneous materials, $\nabla\rho_0=0$, we simplify the notation and use this for the gradient energy density, $W^\nabla = \rho_0 \psi^\nabla$, and use the following decomposition:
\begeq\label{interfacial.energy}
W^\nabla(\eta, \xi, \nabla \eta, \nabla \xi)=W_{1}^{\nabla}(\eta, \xi)+W_{2}^{\nabla}(\xi, \nabla \eta)+W_{3}^{\nabla}(\xi)+W_{4}^{\nabla}(\nabla \xi).
\eqend
The first term consists of a standard double-well potential \cite{levitas2009displacive, levitas2014phase, levitas2013phase}
\begin{eqnarray}
W_{1}^{\nabla}(\eta, \xi)=A\eta ^{2}\left(1-\eta \right)^{2}\iota(\xi),
\end{eqnarray}
where $A=\sfrac{12\Gamma}{l}$ characterizes the energy barrier between two stable phases (minima), relating to the equilibrium energy per unit area, $\Gamma$, and thickness, $l$, of an unstressed interface \cite{clayton2011phase}; $\iota(\xi)$ is a coupling degradation function which degrades with the fracture parameter $\xi$. It is assumed that $\iota(\xi)=g(\xi)$, meaning that the twin boundary energy and the elastic deformation energy degrade with damage according to the same quadratic function. The regularization length is taken as the cohesive process zone for shear failure \cite{clayton2016finsler}
\begin{eqnarray}
l=\frac{16\pi\Upsilon}{\mu_0\left(1-\nu_0 \right)},
\end{eqnarray}
where $\Upsilon$ is the fracture surface energy, $\sfrac{\mu_0}{2\pi}$ is the theoretical shear failure strength, and $\nu_0=\sfrac{\left(3k_0-2\mu_0\right)}{\left(6k_0+2\mu_0\right)}$ \cite{greaves2011poisson}. The second term on the right-hand side of Eq.\,\eqref{interfacial.energy} follows from the Cahn--Hilliard formalism \cite{cahn1958free}
\begin{eqnarray}
W_{2}^{\nabla}(\xi, \nabla \eta)=\kappa_{ij} \eta_{,i}\eta_{,j},
\end{eqnarray}
where $\kappa_{ij}=\kappa_{0}\iota(\xi) \delta_{ij}$ is a diagonal tensor of rank two, and $\kappa_{0}=\sfrac{3\Gamma l}{4}$ is a gradient energy parameter. For cleavage fracture, which is the primary failure mode in boron carbide, we choose the terms in Eq.\,\eqref{interfacial.energy} as follows
\begal
W_{3}^{\nabla}(\xi) =& \mathcal{B}\xi^{2},\\
W_{4}^{\nabla}(\nabla \xi) =& \omega_{ij}\xi_{,i}\xi_{,j} , \ 
\omega_{ij} = \omega_{0}\big(  \delta_{ij}+\beta (\delta_{ij}-M_{i}M_{j} ) \big) ,
\alend
where $\mathcal{B}=\sfrac{\Upsilon}{h}$ is the ratio of fracture surface energy and crack thickness, $\omega_{0}=\Upsilon h$ is a material constant, $\beta$ is the cleavage anisotropy factor, and $\boldsymbol M$ is a unit vector in material coordinates that is normal the cleavage plane \cite{miehe2010thermodynamically, miehe2010phase}. The cleavage plane can be a plane of low surface energy or low intrinsic strength in the crystal \cite{oleinik1996effect}. Generally, there is no predefined relation between cleavage and twinning planes. Orientations of these planes are specified \textit{a priori} and may or may not coincide \cite{clayton2015phase}. The parameter $\beta$ penalizes fracture on planes not normal to $\boldsymbol M$ so that $\beta=0$ results in isotropic damage. This formulation has been used in recent continuum models of fracture as a result of its ability to converge to the correct surface energy of a singular surface when the twin boundary thickness tends to zero \cite{amor2009regularized, del2007variational}. 

By using the aforementioned material modeling and strain definition in Eq.\,\eqref{strain.def}, the governing equations \eqref{gov.eq1} read for displacement
\begeq
P_{ji,j} = 0 \, , \ 
P_{ji} = g(\xi) \pd{W^\text{M}}{F_{ij}} 
= g(\xi) \pd{W^\text{M}}{E^\text{E}_{kl}} \pd{E^\text{E}_{kl}}{F_{ij}} 
= g(\xi) S_{kl} F^\text{E}_{il} (\ten F^\eta)^{-1}_{jk}
\eqend
since
\begal
\pd{W^\text{M}}{E^\text{E}_{kl}} =& S_{kl} = C_{klij} E^\text{E}_{ij} \ , \ S_{kl}=S_{lk} \ ,
\\
\pd{E^\text{E}_{kl}}{F_{ij}} 
=& \pd{E^\text{E}_{kl}}{F^\text{E}_{mn}} \pd{F^\text{E}_{mn}}{F_{ij}} 
= \frac12 \pd{}{F^\text{E}_{mn}} \Big( F^\text{E}_{ok} F^\text{E}_{ol} -\delta_{kl} \Big)
\pd{}{F_{ij}} \Big( (\ten F^\eta)^{-1}_{pn} F_{mp} \Big)
\\
=&\frac12 \Big( \delta_{om} \delta_{kn} F^\text{E}_{ol} + F^\text{E}_{ok} \delta_{om} \delta_{ln} \Big) (\ten F^\eta)^{-1}_{pn} \delta_{mi} \delta_{pj}
=\frac12 \Big( F^\text{E}_{il} (\ten F^\eta)^{-1}_{jk} + F^\text{E}_{ik} (\ten F^\eta)^{-1}_{jl} \Big) \ .
\alend
For phase-fields, in the case of a homogeneous material $\rho_0=\text{const.}$
\begal
 \frac{1}{\mathcal L^\eta} \Dt\eta =&
 - \rho_0 \pd{\psi}{\eta} 
 + \rho_0 \Big( \pd{\psi}{\eta_{,i}}\Big)_{,i} 
 = 
 - g(\xi)\pd{W^\text{M}}{\eta} 
 - \pd{W_{1}^{\nabla}}{\eta} 
 + \Big( \pd{ W_{2}^{\nabla} }{\eta_{,i}}\Big)_{,i} 
 \\
 =& 
 - \frac12 g(\xi) E^\text{E}_{ij} \big( C^\text{T}_{ijkl} - C^\text{P}_{ijkl} \big) \phi'(\eta) E^\text{E}_{kl} 
 - g(\xi) P_{ji} \pd{F_{ij}}{\eta} 
 \\
 &- \Big( 2 A \eta (1-\eta)^2  - A\eta^2 2(1-\eta)  \Big) g(\xi) 
 + \Big( 2 \kappa_0 g(\xi) \eta_{,i} \Big)_{,i}
 \ ,
\alend
and
\begal
 \frac{1}{\mathcal L^\xi}  \Dt\xi =&
 - \rho_0 \pd{\psi}{\xi} 
 + \rho_0 \Big( \pd{\psi}{\xi_{,i}}\Big)_{,i} 
 =
 - g'(\xi) W^\text{M} 
 -  \pd{(W_{1}^{\nabla}+W_{2}^{\nabla}+W_{3}^{\nabla})}{\xi} 
 + \Big( \pd{W_{4}^{\nabla}}{\xi_{,i}}\Big)_{,i} 
 \\
 =& 
 - \frac12 g'(\xi) E^\text{E}_{ij} C_{ijkl} E^\text{E}_{kl}
 - A \eta^2 (1-\eta)^2 g'(\xi) 
 - \kappa_0 g'(\xi) \eta_{,i} \eta_{,i} 
 - 2 \mathcal{B} \xi
 \\
 &+ \Big( 2 \omega_{0}\big(  \delta_{ij}+\beta (\delta_{ij}-M_{i}M_{j} ) \big) \xi_{,j} \Big)_{,i}
 \ .
\alend
By using Eqs.\,\eqref{eq.1}, \eqref{strain.def}, we have
\begeq
\pd{F_{ij}}{\eta} = F^\text{E}_{ik} \phi'(\eta) \gamma_0 s_k m_j
\eqend
For a better analogy, we misused the notation for the same degradation function in Eq.\,\eqref{degradation} as follows:
\begeq
g(\xi)=\zeta + (1-\zeta) \big( 1-\xi \big)^2 , \ 
g'(\xi) = - 2(1-\zeta) \big(1-\xi \big).
\eqend
Finally, we obtain the following equations to solve numerically,
\begeq
 \frac{1}{\mathcal L^\eta} \Dt\eta 
 + \frac12 g(\xi) E^\text{E}_{ij} \big( C^\text{T}_{ijkl} - C^\text{P}_{ijkl} \big) \phi'(\eta) E^\text{E}_{kl} 
 + g(\xi) P_{ji} F^\text{E}_{ik} \phi'(\eta) \gamma_0 s_k m_j 
 \\
 + 2 A \big( \eta -3\eta^2 + 2\eta^3 \big) g(\xi) 
 - \Big( 2 \kappa_0 g(\xi) \eta_{,i} \Big)_{,i}
 =0
 \ ,
\eqend
and
\begeq
 \frac{1}{\mathcal L^\xi}  \Dt\xi 
 + \frac12 g'(\xi) E^\text{E}_{ij} C_{ijkl} E^\text{E}_{kl}
 + A \eta^2 (1-\eta)^2 g'(\xi) 
 + \kappa_0 g'(\xi) \eta_{,i} \eta_{,i} 
 + 2 \mathcal{B} \xi
 \\
 - \Big( 2 \omega_{0}\big(  \delta_{ij}+\beta (\delta_{ij}-M_{i}M_{j} ) \big) \xi_{,j} \Big)_{,i} 
 =0 \ .
\eqend

\subsection{Variational formulation} \label{section.computation}

We follow the standard techniques for generating weak forms to solve numerically by means of the finite element method \cite{zohdi2018finite}. The space discretization is incorporated by approximating fields, $\ten u$, $\eta$, $\xi$, by spanning over nodal values after a triangulation of the computational domain, $\Omega$, with its closure, $\p\Omega$, into finite elements. For simplicity, we skip a notational change for approximated fields, since their analytical and discrete representations never occur in the same formulation. We emphasize that all unknowns, $\big\{ \ten u, \eta, \xi\big\}$, are solved in a monolithic manner, therefore, the Hilbertian Sobolev space, $\mathscr{H}^n$, with the polynomial order, $n$, as follows:
\begeq
\mathscr{V} =\Bigg\{ \big\{ \ten u, \eta, \xi\big\} \in [ \mathscr{H}^{n}(\Omega) ]^\text{DOF} : \big\{ \ten u, \eta, \xi \big\} \Big|_{\p\Omega} = \text{given}  \Bigg\} \ .
\eqend
We use mixed spaces for quantities, depending on the problem, displacement (linear or quadratic) and phase-fields (linear) standard Lagrange finite element. On each node, unknowns read $2+1+1=4$ degrees of freedom (DOF) in 2-D and $3+1+1=5$ (DOF) in 3-D space. As usual in the Galerkin approach, we use the same space for test functions, $\big\{ \del\ten u, \del\eta, \del\xi\big\}$, where they vanish on Dirichlet boundaries
\begeq
\bar{\mathscr V} =\Bigg\{ \big\{ \del\ten u, \del\eta, \del\xi \big\} \in [ \mathscr{H}^{1}(\Omega) ]^\text{DOF} : \big\{ \del\ten u, \del\eta, \del\xi \big\} \Big|_{\p\Omega} = 0  \Bigg\} \ .
\eqend
To ensure the irreversibility of the order parameter rates in our simulations, $\dot{\eta}$ and $\dot{\xi}$, we restrict them to be zero if they become negative. For the time discretization, we use Euler backwards scheme for order parameters, for example
\begeq
\Dt\eta = \frac{\eta - \eta^0}{\Delta t} \ ,
\eqend
where $\Delta t$ is the time step. For simplicity we use constant time steps. This method is implicit, hence for real valued problems stable, and converges to the correct solution. Multiplying governing equations by test functions, generating integral forms, and then integrating by parts where necessary, we obtain
\begal
\text{Form}_{\ten u} =& -\int_\Omega P_{ji} \del u_{i,j} \dd V + \int_{\p\Omega_\text{N}} \hat t_i \del u_i \dd A \ , 
\alend
where a traction vector, $\hat{\ten t}$, is given on Neumann boundaries, $\p\Omega_\text{N}$.
\begal
\text{Form}_{\eta} =& \int_\Omega \Bigg( 
\frac{\eta - \eta^0}{ \mathcal L^\eta \Delta t} \del \eta
+ \frac12 g(\xi) E^\text{E}_{ij} \big( C^\text{T}_{ijkl} - C^\text{P}_{ijkl} \big) \phi'(\eta) E^\text{E}_{kl} \del\eta
+ g(\xi) P_{ji} F^\text{E}_{ik} \phi'(\eta) \gamma_0 s_k m_j \del\eta
\\
&+ 2 A \big( \eta -3\eta^2 + 2\eta^3 \big) g(\xi) \del\eta
+ 2 \kappa_0 g(\xi) \eta_{,i} \del\eta_{,i}
\Bigg) \dd V,
\alend
where we have employed the fact that $N_i \eta_{,i}$ vanishes at the boundary of $\Omega$ in connection with Eq.\,\eqref{second.law3}. Analogously, we obtain
\begal
\text{Form}_{\xi} =& \int_\Omega \Bigg( 
\frac{\xi - \xi^0}{ \mathcal L^\xi \Delta t} \del\xi
 + \frac12 g'(\xi) E^\text{E}_{ij} C_{ijkl} E^\text{E}_{kl} \del\xi
 + A \eta^2 (1-\eta)^2 g'(\xi) \del\xi
 \\
 &+ \kappa_0 g'(\xi) \eta_{,i} \eta_{,i} \del\xi
 + 2 \mathcal{B} \xi \del\xi
 +  2 \omega_{0}\big(  \delta_{ij}+\beta (\delta_{ij}-M_{i}M_{j} ) \big) \xi_{,j} \del\xi_{,i} 
 \Bigg) \dd V
\alend
The implementation solves the nonlinear weak form
\begeq\label{weak.forms}
\text{Form} = \text{Form}_{\ten u} + \text{Form}_{\eta} + \text{Form}_{\xi},
\eqend
after a symbolic derivation and Newton--Raphson iterations.

\section{Multiphysics simulation} \label{section.results}

The weak forms in Eq.\,\eqref{weak.forms} are nonlinear and coupled. We have implemented a transient, fully coupled solution strategy by using open-source packages from the FEniCS Project \cite{logg2010dolfin}. We refer to \cite{074,barchiesi2021computation,tangella2022hybrid} for implementations in FEniCS by means of a staggered scheme. Herein, the implementation uses a monolithic approach, where displacement and phase fields are solved at once. Staggered solution solves many smaller problems than one larger, which is faster since the computational cost increases exponentially. However, in a staggered algorithm, several iterations are necessary for solving one time step in order to ensure that coupling between unknowns are fulfilled. Generally speaking, for highly-coupled systems, a monolithic approach is more feasible. For the linearization, we use a standard Newton--Raphson approach. The linearization is done automatically by means of a symbolic derivation that allows the user to write the weak form without going through the error-prone linearization process by hand \cite{olgaard2009automated,olgaard2010optimizations}. The code is written in Python, although the FEniCS software wraps the formulation to a C\texttt{++} code and solves as a compiled program. Therefore, yet efficient in developing the code, all computation is running in parallel very efficiently. In short, the problem-specific parts of the computer code used to perform the simulations have been generated automatically from a high-level description that resembles closely the notation used in this work.

Examples under different loading conditions in two-dimensional samples are demonstrated next in order to simulate deformation mechanisms observed in metallic magnesium and ceramic boron carbide. The results are adequate, qualitatively and quantitatively. The material properties used in the simulations are shown in Table \ref{accuracy} for \ch{Mg} and \ch{B4C}. Five independent second-order elastic constants \cite{hill1952elastic} are listed by using Voigt notation, with indices $I$, $J$ from 1 to 6, as follows:
\begeq
C_{IJ}=
\begin{pmatrix}
C_{1111} & C_{1122} & C_{1133} & C_{1123} & C_{1113} & C_{1112} \\
 & C_{2222} & C_{2233} & C_{2223} & C_{2213} & C_{2212} \\
&  & C_{3333} & C_{3323} & C_{3313} & C_{3312} \\
 & & & C_{2323} & C_{2313} & C_{2312} \\
& \text{sym.} &  &  & C_{1313} & C_{1312} \\
 &  &  &  &  & C_{1212}
\end{pmatrix} \ ,
\eqend
The bulk modulus listed below for each undamaged material is obtained by \cite{clayton2014analysis}
\begeq\label{elastic.constants}
k = \frac{\left(C_{11} + C_{12} \right)C_{33} - 2C_{13}^{2}}{C_{11}+C_{12}+2C_{33}-4C_{13}}
\eqend
%
%
\begin{table}[htb]
\centering 
\caption{Material properties and model constants for magnesium and boron carbide} \vspace{.2cm} \label{accuracy}
\setlength{\tabcolsep}{0.8em}
\renewcommand{\arraystretch}{1.4}
\begin{tabular}{c r l l c c c c}
Parameters & Notation & Value-\ch{Mg} & Value-\ch{B4C} & Reference  \\
\hline
\multirow{5}{*}{Elastic constants} & $C_{11}=$ & $\unit[63.5]{GPa}$ & $\unit[487]{GPa}$ & \multirow{5}{*}{\cite{PhysRev.107.972, taylor2011first}}        \\
                                   & $C_{12}=$ & $\unit[25.9]{GPa}$ & $\unit[117]{GPa}$ & \\
                                   & $C_{13}=$ & $\unit[21.7]{GPa}$ & $\unit[66]{GPa}$ & \\
                                   & $C_{33}=$ & $\unit[66.5]{GPa}$ & $\unit[525]{GPa}$ & \\
                                   & $C_{44}=$ & $\unit[18.4]{GPa}$ & $\unit[133]{GPa}$ & \\
Shear modulus & $\mu=$    & $\unit[19.4]{GPa}$ & $\unit[193]{GPa}$ & \cite{clayton2012towards, wang20091}        \\
Bulk modulus & $k=$  & $\unit[36.9]{GPa}$  & $\unit[237]{GPa}$ &    Eq.  (48)  \\
Twin surface energy & $\Gamma=$ & $\unit[0.12]{\sfrac{J}{m^2}}$ & $\unit[0.54]{\sfrac{J}{m^2}}$ &  \cite{hirth1983theory}    \\
Fracture surface energy & $\Upsilon=$ & $\unit[7.26]{\sfrac{J}{m^{2}}}$  & $\unit[3.27]{\sfrac{J}{m^{2}}}$ &  \cite{clayton2013phase, beaudet2015surface}    \\
Twinning shear & $\gamma_0=$  & $0.13$ & $0.31$ & \cite{li2010deformation, christian1995deformation}       \\
Gradient energy parameter & $\kappa_0=$  & $\unit[0.0878]{\sfrac{nJ}{m}}$ & $\unit[0.4212]{\sfrac{nJ}{m}}$ &  Eq. (30)    \\
Transformation barrier & $A=$ & $\unit[1.404]{GPa}$ & $\unit[3.01]{GPa}$ & \cite{fanchini2006behavior, christian1995deformation}         \\
Regularization length & $l = h =$ & $\unit[1.00]{nm}$ & $\unit[1.04]{nm}$ & Eq. (29)       \\
Kinetic coefficient (Twinning) & $\mathcal{L}^{\eta}=$ & $\unit[4200]{\left(Pa \cdot s \right)^{-1}}$ & $\unit[2000]{\left(Pa \cdot s \right)^{-1}}$ & \\
Kinetic coefficient (Fracture) & $\mathcal{L}^{\xi}=$ & suppressed $(\xi = 0)$ & $\unit[1000]{\left(Pa \cdot s \right)^{-1}}$ & \\
\hline
\end{tabular}
\label{tab.2}
\end{table}
According to the primary inelastic mechanisms for each material mentioned before, three different problems are simulated and discussed in the following to represent the degenerate cases:
\begin{enumerate}[(i)]
\item Twin propagation in two-dimensional single crystals magnesium and boron carbide in Sect.\,\ref{simulation1}; Fracture is suppressed in this condition by assuming $\xi= 0$. 
\item Analysis of twinning induced by a crack in magnesium under pure mode \Romannum{1} or mode \Romannum{2} loading in Sect.\,\ref{simulation2}; Similar to the previous case, fracture is not calculated, $\xi = 0$.
\item Fracture in homogeneous single crystal boron carbide under biaxial compressive loading in Sect.\,\ref{simulation3}; Twinning is suppressed for this problem by setting $\eta=0$.
\end{enumerate}
These examples demonstrate that we generate knowledge about mechanical deformations in very small length-scales and extreme loading rates causing a twin or crack initiation and propagation at very small time scales. These extreme conditions are challenging to observe experimentally, where we rely on accurate multiphysics simulations as presented herein.

\subsection{Twin growth and propagation}\label{simulation1}

\begin{figure}[H]
\centering
\includegraphics[width=70mm, height=70mm]{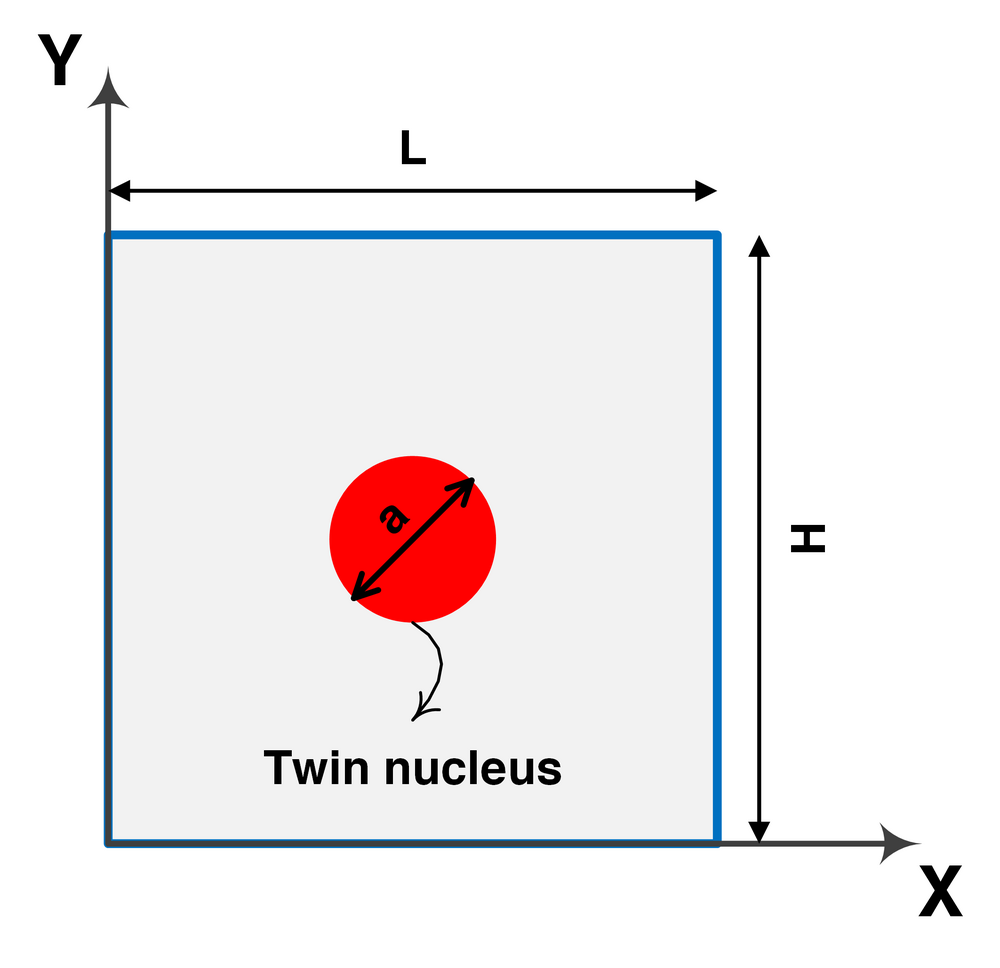}
\caption{The numerical setup of the rectangular single crystal (gray), $\eta=0$, including a single twin embryo (red), $\eta=1$.}
\label{fig.2}
\end{figure}

Nucleation and evolution of deformation twinning in a single crystal of magnesium (in Sect.\,\ref{simulation1.1}) and boron carbide (in Sect.\,\ref{simulation1.2}) are presented in a two-dimensional domain in plain strain conditions to be depicted in Fig.\,\ref{fig.2}.
\noindent For validation, the model is initially solved for elastically isotropic pure magnesium single crystals with the properties listed in Table 1. The isotropic elastic approximation appears reasonable because magnesium single crystals are not strongly anisotropic elastically \cite{hearmon1946elastic}.
A circular twin nucleus, $\eta=1$, of initial radius $a=\unit[3]{nm}$ is embedded in a rectangular domain with a surrounding parent material, $\eta=0$. The domain is of $\unit[40\times40]{nm}$ in size for the magnesium simulations. The initial radius of the twin embryo is set to $\unit[3]{nm}$ as a result of the fact that a bifurcation from circular to elliptical shape occurs for a radius of $\unit[3.2]{nm}$, corresponding to the analytical sharp interface solution \cite{lee1990elastic}. The lattice orientation vectors are in the form
\begin{eqnarray}
\ten s=\begin{pmatrix} \cos(\theta) \\ \sin(\theta) \end{pmatrix} , \  
\ten m=\begin{pmatrix} -\sin(\theta) \\ \cos(\theta) \end{pmatrix} ,
\end{eqnarray}
where $\theta$ denotes the orientation of the habit plane. Also, according to the following matrix-form gradient coefficient 
\begin{eqnarray}
\boldsymbol \kappa=
\begin{pmatrix}
\kappa_{11} & 0 \\
0 & \kappa_{22} \\
\end{pmatrix},
\end{eqnarray}
both isotropic ($\kappa_{11}=\kappa_{22}=\kappa_0$) and anisotropic ($\frac{\kappa_{11}}{2}=2\kappa_{22}=\kappa_0$) twin boundary surface energies are employed in different simulations in order to explore their effects as well as for validation purposes. The following simple shear with Dirichlet boundary conditions on $\partial \mathfrak{B}_\text{D}$ for top and bottom boundaries are used
\begin{eqnarray}
\ten u\Big|_{_\text{D}} = \begin{pmatrix} \Lambda Y \\ 0 \end{pmatrix} , 
\eta\Big|_{\partial \mathfrak{B}_\text{D}} = 0 ,
\end{eqnarray} 
where $\Lambda=0.08$ is the magnitude of applied shear for all simulations in the following section. The twin growth to the boundary is inhibited by the displacement boundary conditions. The order parameter gradients also vanish at the boundaries due to the Neumann boundary conditions defined in Eq.\,\eqref{boundary}.

\subsubsection{Twin embryo propagation and growth in single crystal magnesium}\label{simulation1.1}

Figure\,\ref{fig.3} shows contour plots demonstrating the spatial distributions of numerical results for the growth of a circular twin embryo in a single crystal magnesium with an orientation of the habit plane $\theta = 0$. The embryo is undergoing a simple shear at 8\% displacement prescribed on the top. Parameters of interest include the twin order parameter (i, ii), \textit{y} displacement (iii, iv), and shear stress (v, vi). Each image pair considers both small (left side) and large strains (right side), as well as isotropic (a, b) and anisotropic surface energies (c, d). For this case, there is no significant difference in the simulation results between linear (left side) and nonlinear (right side) elasticity. The results are shown at time instants of $t=\unit[50]{ps}$ and $t=\unit[500]{ps}$ to show the evolution of the twin’s morphology. The mesh of the rectangular domain includes 160,000 linear triangular elements. By using a standard $h$-convergence, we have chosen this particular mesh to deliver mesh insensitive results. The \hkl <1 0 -1 1> plane and \hkl {-1 0 1 2} direction are considered as the primary twinning system in magnesium \cite{STAROSELSKY20031843}.

First, the evolution of the twin order parameter is shown in Fig.\,\ref{fig.3}(a, b)(i, ii) under simple shear with the boundary conditions defined at $t=\unit[50]{ps}$ and $t=\unit[500]{ps}$ for small and large deformation with isotropic twin boundary energy. As can be seen, the twin embryo grows until it is repelled by the rigid outer boundaries, where the order parameter is set to zero. Under these numerical conditions, a small orientation of the twin evolution is realized due to the difference in the driving force for twinning, which is a factor of $\left(\boldsymbol F^{\eta}\right)^{-1}$. We emphasize that the twin morphology at the final stage is in qualitative agreement with the (static) phase-field results \cite{clayton2011phase} and molecular dynamics simulations \cite{hu2020embracing}, thus serving to verify the set of results in Fig.\,\ref{fig.3}(a, b)(i, ii) for the proposed time-dependent phase-field model.

Second, the distribution of the displacement in the \textit{y} direction for the domain under simple shear loading for small and large strains at different times are depicted in Fig.\,\ref{fig.3}(a, b)(iii, iv). The positive and negative displacement values indicate that the left and right sides of the twinned boundary regions are under compressive and tensile loading, respectively. This distribution in a simple shear is not possible for a homogeneous material with prescribed vanishing $y$ displacement on top. Herein, we stress that the parent to twin phase change introduces a heterogeneity in stiffness parameters across the twin boundaries. By means of this relatively simple simulation, we gather an insight into the material response. Moreover, the range of displacement magnitudes at the very last time instant are lower than those at initial times as a result of inhibiting by the boundaries. The corresponding evolution of the shear stress for small and large strains with consideration of the isotropic surface energy at various times are illustrated in Fig.\,\ref{fig.3}(a, b)(v, vi). Investigating the shear stress distribution improves our knowledge of the redistribution of high local stress, resulting from twinning \cite{CLAUSEN20082456}, and this provides new insights into demonstrating the driving force for the propagation and growth of twin within a small region in the microstructure.
\newgeometry{margin=0.5cm, includefoot}
\begin{landscape}
\thispagestyle{mylandscape} 

\begin{figure}[h]
\centering
\includegraphics[width=242.90mm, height=151.88mm]{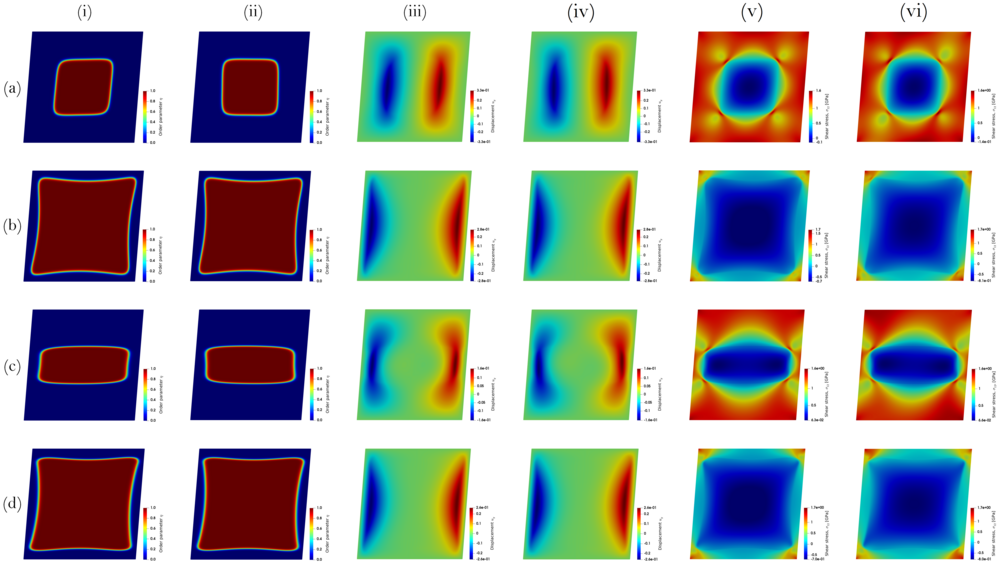}
\caption{Evolution of a circular twin embryo in a rectangular domain in single crystal magnesium in both small (left side in image pair) and large (right side in image pair) deformations considering isotropic and anisotropic surface energies and elasticity with orientation of the habit plane $\theta=0$: (a) row shows isotropic surface energy case at $t=\unit[50]{ps}$, (b) row shows isotropic surface energy case at $t=\unit[500]{ps}$, (c) row shows anisotropic surface energy case at $t=\unit[50]{ps}$, and (d) row shows anisotropic surface energy case at $t=\unit[500]{ps}$. (i) column shows twin order parameter contour for small deformation, (ii) column shows twin order parameter contour for large deformation, (iii) column shows displacement in \textit{y} direction contour for small deformation, (iv) column shows displacement in \textit{y} direction contour for large deformation, (v) column shows shear stress contour for small deformation, and (v) column shows shear stress contour for large deformation. (For interpretation of the references to color in this figure, the reader is referred to the web version of this article.)}
\label{fig.3}
\end{figure}

\end{landscape}
\restoregeometry
Third, in Fig.\,\ref{fig.3}(a, b)(v, vi), one component of the stress tensor (shear) is shown. Again, in a homogeneous material under simple shear conditions, a constant shear stress is created. The distribution is caused by the twinning, as visible by the shape compared to the twin distribution. 

Fourth, the effect of anisotropic surface energy is studied in Fig.\,\ref{fig.3}(c, d)(i - vi). For the twin order parameter in Fig.\,\ref{fig.3}(c)(i, ii), the equilibrium shape of the twin embryo under small strains is wider in the horizontal direction (parallel to the habit plane) and flatter in the vertical direction at $t=\unit[50]{ps}$ as compared to the isotropic energy case shown in Fig.\,\ref{fig.3}(a)(i, ii). This behavior has been observed previously in the time-independent phase-field approach \cite{clayton2011phase}, where the results are in agreement with those in this study, thus providing an additional confirmation of this phenomenon. After completing its growth in the horizontal direction, Fig.\,\ref{fig.3}(c)(i, ii), the twin begins to grow in width for later times, $t=\unit[500]{ps}$, Fig.\,\ref{fig.3}(d)(i, ii). This behavior is correlated to the surface energy anisotropy ratio $\sfrac{\kappa_{11}}{\kappa_{22}}$. Moreover, the twin interface thickness has a lower value in the direction normal to the habit plane for the anisotropic surface energy scenario depicted in Fig.\,\ref{fig.3}(c, d)(i, ii) as compared with the isotropic case from Fig.\,\ref{fig.3}(a, b)(i, ii). This phenomenon is related to the contribution of the core and elastic energies to the total surface energy of the interface \cite{Kosevich_1971}. The displacement for the anisotropic case in Fig.\,\ref{fig.3}(c, d)(iii, iv) is lower than in the isotropic one in Fig.\,\ref{fig.3}(a, b)(iii, iv). Finally, the variation of shear stresses for anisotropic surface energies at various time instants under small and large strains are depicted in Fig.\,\ref{fig.3}(c, d)(v, vi). Considering the results at $t=\unit[500]{ps}$, Fig.\,\ref{fig.3}(c, d)(vi), the maximum and minimum shear stress values for the current simulations are within a 7\% difference of the results obtained in \cite{clayton2011phase} by means of a static simulation, demonstrating the significance of inertial terms in extreme loading conditions. For both isotropic and anisotropic surface energies, the magnitude of the shear stress within the twinning region decreases as a function in time and, eventually, becomes negative. This observation is consistent with experimental results for single crystal magnesium under simple shear loading \cite{ArulKumar2018}.

For the next set of simulation examples in Fig.\,\ref{fig.4}, the same boundary conditions and numerical setup from Fig.\,\ref{fig.3} are considered for $\theta = \sfrac{\pi}{6}$. The layout of the figure is similar to that of Fig.\,\ref{fig.3} with $\theta = 0$ where (a, b)(i - vi) and (c, d)(i - vi) are the simulation results for the order parameter, displacement, and shear stress under small and large strains at $\unit[t = 50]{ps}$ and $\unit[t = 500]{ps}$ for isotropic and anisotropic surface energies, respectively. For the isotropic surface energy case in Fig.\,\ref{fig.4}(a)(i, ii) at $\unit[t = 50]{ps}$, the twin is smaller as a consequence of less driving force under the same shear loading of 8\% as compared with Fig.\,\ref{fig.3}(a)(i, ii). Further, the twin area fraction at $\unit[t = 500]{ps}$ shown in Fig.\,\ref{fig.4}(b)(i, ii) is much smaller than the case, when the orientation of the habit plane is aligned with the shear loading direction (previously in Fig.\,\ref{fig.3}(a)(i, ii)). In the case of large strains, the twin tends to grow more prominently in the direction of the habit plane when $\theta = \sfrac{\pi}{6}$ than when $\theta = 0$ (Fig.\,\ref{fig.4}(a)(ii)). The displacement contours shown in Fig.\,\ref{fig.4}(a - d)(iii, iv) indicate that the upper and lower sides of the twin’s interface are under tensile and compressive loading, respectively, which is similar to Fig.\,\ref{fig.3}(a - d)(iii, iv). The displacement in the vertical direction (Fig.\,\ref{fig.4}(a)(iv)) is $\sim$17\% greater than that for the small deformation case depicted in Fig.\,\ref{fig.4}(a)(iii), and the maximum shear stress under large strain conditions (Fig.\,\ref{fig.4}(a)(vi)) is 5\% greater than that for the small deformation case (Fig.\,\ref{fig.4}(a)(v)). At $\unit[t = 500]{ps}$, the twin embryo has a greater thickness for small deformations (Fig.\,\ref{fig.4}(b)(i)) as compared with its growth in length in the direction of the habit plane for the case of large deformations (Fig.\,\ref{fig.4}(b)(ii)), until it is prohibited by the boundaries. The displacement at the end of the simulation is around 17\% larger for small strains (Fig.\,\ref{fig.4}(b)(iii)) as compared with the large deformation result (Fig.\,\ref{fig.4}(b)(iv)). Lastly, the spatial variations of shear stress at $t = \unit[50]{ps}$ and $t = \unit[500]{ps}$ are depicted in Fig.\,\ref{fig.4}(a, b)(v, vi). As can be seen, the minimum and maximum shear stress values happen in the twinned region and matrix, respectively. The heterogeneous stress distribution around the twins is due to a sudden change in the stresses within the twin interface \cite{LIU2018203}.

Next, the phase-field results for the anisotropic surface energy and $\theta = \sfrac{\pi}{6}$ are shown in Fig.\,\ref{fig.4}(c, d)(i-vi). Considering the distribution of the twin order parameter for small strains, Fig.\,\ref{fig.4}(c, d)(i, ii), the twin boundaries tend to be expanded parallel to the habit plane when compared with the isotropic case because the elongation in the direction of $\boldsymbol s$ is favored due to a decreasing contribution of the gradient energy term \cite{clayton2011phase}. Pointing to Fig.\,\ref{fig.4}(c)(iii, iv), the maximum displacement values for large deformations are 20\% higher than those in the small deformation case from Fig.\,\ref{fig.4}(a)(iii, iv). At the tip of the twin, the shear stress is maximum and $\sim$10\% larger for large strain conditions (Fig.\,\ref{fig.4}(a)(vi)) as compared to the small deformation case (Fig.\,\ref{fig.4}(a)(v)). For the same boundary conditions, the results are depicted for $t=\unit[500]{ps}$ in Fig.\,\ref{fig.4}(b, d). Here, the twin embryo has a different equilibrium shape than what was shown in Fig.\,\ref{fig.3} for $\theta=0$. Namely, the twin is rotated in such a way that one axis in the reference coordinate is aligned to the direction $\boldsymbol{s}$ of twinning shear, as shown in Fig.\,\ref{fig.4}(d)(i, ii). The twin interface also has a lower thickness in the direction normal to the habit plane due to the various contributions of the core and elastic energies to the interface energy \cite{Kosevich_1971}. For the displacement contour, the values are 30\% larger for the anisotropic energy (Fig.\,\ref{fig.4}(d)(iii, iv)) as compared to the isotropic case, while the difference in shear stress for small and large strains is negligible.
\newgeometry{margin=0.5cm, includefoot}
\begin{landscape}
\thispagestyle{mylandscape} 
\begin{figure}[h]
\centering
\includegraphics[width=252.90mm, height=151.88mm]{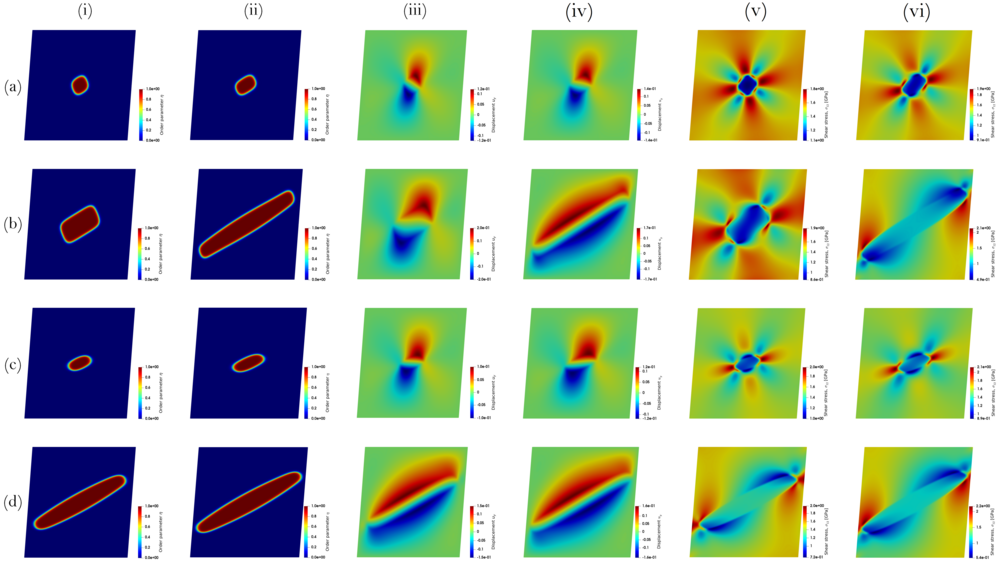}
\caption{Evolution of a circular twin embryo in a rectangular domain in single crystal magnesium in both small (left side in image pair) and large (right side in image pair) deformations considering isotropic and anisotropic surface energies and elasticity with orientation of the habit plane $\theta=\sfrac{\pi}{6}$: (a) row shows isotropic surface energy case at $t=\unit[50]{ps}$, (b) row shows isotropic surface energy case at $t=\unit[500]{ps}$, (c) row shows anisotropic surface energy case at $t=\unit[50]{ps}$, and (d) row shows anisotropic surface energy case at $t=\unit[500]{ps}$. (i) column shows twin order parameter contour for small deformation, (ii) column shows twin order parameter contour for large deformation, (iii) column shows displacement in \textit{y} direction contour for small deformation, (iv) column shows displacement in \textit{y} direction contour for large deformation, (v) column shows shear stress contour for small deformation, and (v) column shows shear stress contour for large deformation. (For interpretation of the references to color in this figure, the reader is referred to the web version of this article.)}
\label{fig.4}
\end{figure}
\end{landscape}
\restoregeometry

\subsubsection{Twin embryo propagation and growth in single crystal boron carbide}\label{simulation1.2}

For the first time in the literature, the numerical results obtained from phase-field approach are validated with the high-resolution transmission electron microscopy (HRTEM) \cite{mackinnon1986high} for the twinning propagation, growth, and interactions in \ch{B4C}. The ubiquitous existence of twins and stacking faults in pressureless sintered and hot-pressed \ch{B4C}, reported in the previous literature \cite{thevenot1990boron, mutsuddy1987mechanical}, has motivated studies of their impact \cite{awasthi2020deformation, zhang2021size}. This is important because it is widely accepted that existing nanotwins, ranging from $t=\unit[1]{nm}$ up to $t=\unit[30]{nm}$ in width for milled and unmilled samples \cite{heian2004synthesis}, would enhance the strength and hardness of boron carbide \cite{an2016superstrength} by arresting twin boundary slip within the nanotwins \cite{an2016superstrength}. As a result, the presented results opens a number of interesting possibilities for simulating and controlling microstructure pattern development in materials experiencing extreme mechanical loading \cite{orowan1954dislocations, hooshmand2019atomic}. Given the lack of true images of the twin interfaces in boron carbide \cite{li2010deformation, anselmi2004modeling} and the difficulty in experimentally tracking the twin growth process, the present continuum mechanics model will provide insight into the deformation behavior of pre-existed twinned \ch{B4C}, which have been largely neglected in previous works \cite{clayton2011phase, clayton2018continuum}. In addition, the morphology of mature twins will be affected by the early stages of the twin nucleus evolution, which necessitates a comprehensive model as herein. In this light, understanding how twins are formed and then developing effective strategies for incorporating twin boundaries into polycrystalline microstructures constitute an attractive approach for enhancing the mechanical response of ceramics. To address this, we conducted numerical simulations using the proposed phase-field model in a boron carbide single crystal. 

\begin{figure}[H]
\centering
\includegraphics[width=0.8\textwidth, height=100mm]{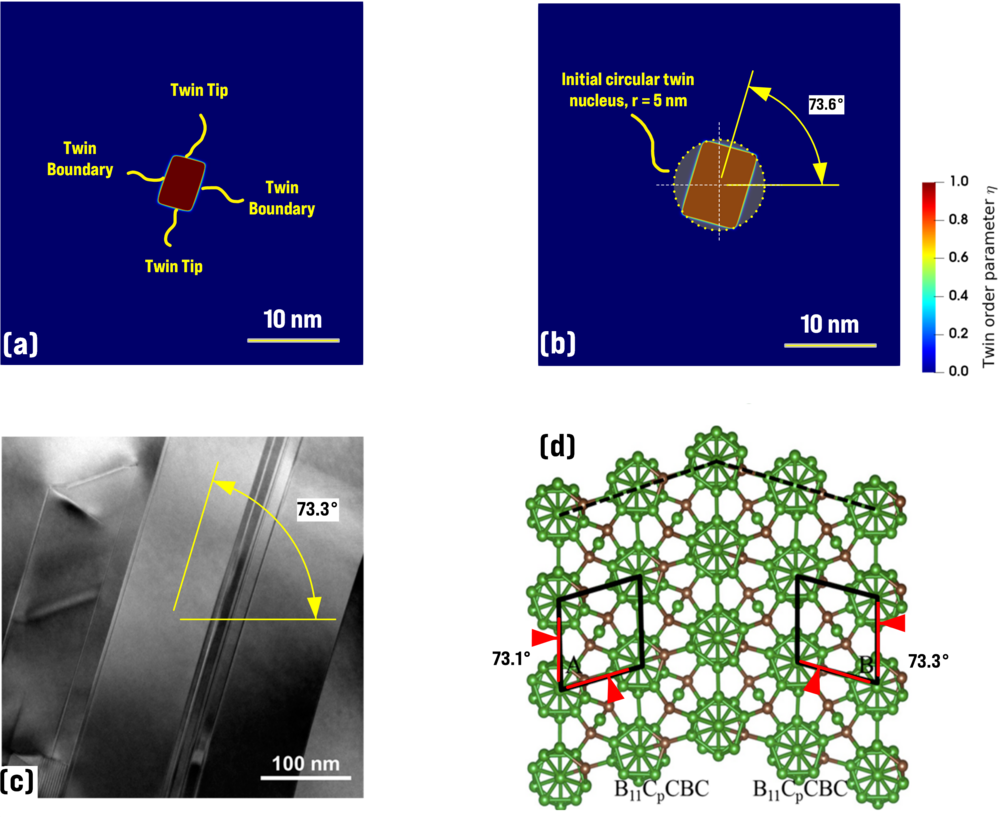}
\caption{The distribution of twin order parameter in boron carbide from snapshots taken at (a) $\unit[1]{ps}$ and (b) $\unit[2]{ps}$ for samples deformed under shear strain; (a) the various interfaces associated with a single twin embryo; (b) the direction of twin propagation in a $\unit[40]{nm} \times \unit[40]{nm}$ plate with an initial circular twin nucleus of radius $\unit[5]{nm}$ denoted by yellow dotted line; (c) the TEM image showing the larger twin spacing in boron carbide at the $\unit[100]{nm}$ scale with an angle of $\unit[73.3^{\circ}]{}$ \cite{an2016superstrength}; and (d) the symmetric twin in boron carbide with inclination angles of $\unit[73.1^{\circ}]{}$ and $\unit[73.3^{\circ}]{}$ on the two sides predicted by density functional theory \cite{an2016superstrength}. (c) and (d) reproduced with permission from \cite{an2016superstrength}. (For interpretation of the references to color in this figure, the reader is referred to the web version of this article.)}
\label{fig.5}
\end{figure}

The combination of growth of a single twin embryo is measured along two critical directions, including twin thickening through twin boundary (TB) migration and twin tip (TT) propagation. The simulation results are then compared with experiments in Fig.\,\ref{fig.5}. Shear strains are applied by displacing all the boundary regions, while the bottom side is fixed. A time step of $\Delta t=\unit[1]{fs}$ is chosen for solving the problem. The dimensions of the simulation domain are $\unit[40]{nm} \times \unit[40]{nm}$ in the \textit{X} and \textit{Y} directions, and contains 160,000 linear triangular elements. One circular twin embryo with a radius of $\unit[5]{nm}$ is inserted at the center of a square containing the perfect \ch{B4C} crystal lattice, using the Eshelby method as in \cite{xu2013importance}. The magnitude of applied shear $\Lambda$ is set to $0.3$, which is maximum at the top and zero at the bottom. Additional simulations showed that choosing a shear magnitude lower than $0.3$ leads to shrinking and disappearing of the twin.
Schematics of the simulation result for an initially circular twin embryo in boron carbide at $t=\unit[1]{ps}$ and $t=\unit[2]{ps}$ with the $X$-axis along the $[1 \overline{1} 0 \overline{1}]$ direction are shown in Figs.\,\ref{fig.5}(a) and \ref{fig.5}(b), respectively. Figure\,\ref{fig.5}(a) depicts the ``twin tip'', which occurs in the primary direction of twin growth, and the ``twin boundary'', which occurs in a direction perpendicular to the twin growth. Under shear loading, the size and shape of the initial circular twin has changed until reaching a stable configuration. Similar to other ceramics such as calcite, the twin was contracted at the beginning of loading, and this has been shown to be related to the stress reversal \cite{doi:10.1063/1.1658166}. Next, the twin embryo's shape and growth direction at $t=\unit[2]{ps}$ from Fig.\,\ref{fig.5}(b) is compared with the high-resolution transmission electron microscope images (Fig.\,\ref{fig.5}(c)) and density functional theory results (Fig.\,\ref{fig.5}(d)) \cite{an2016superstrength}. The shape and angle of the twin embryo obtained from the numerical simulations are in good agreement with the previously published results, showing the symmetric twin with an inclination angle of $73.1^{\circ}$ to $73.3^{\circ}$.

\begin{figure}[H]
\centering
\includegraphics[width=0.95\textwidth, height=120mm]{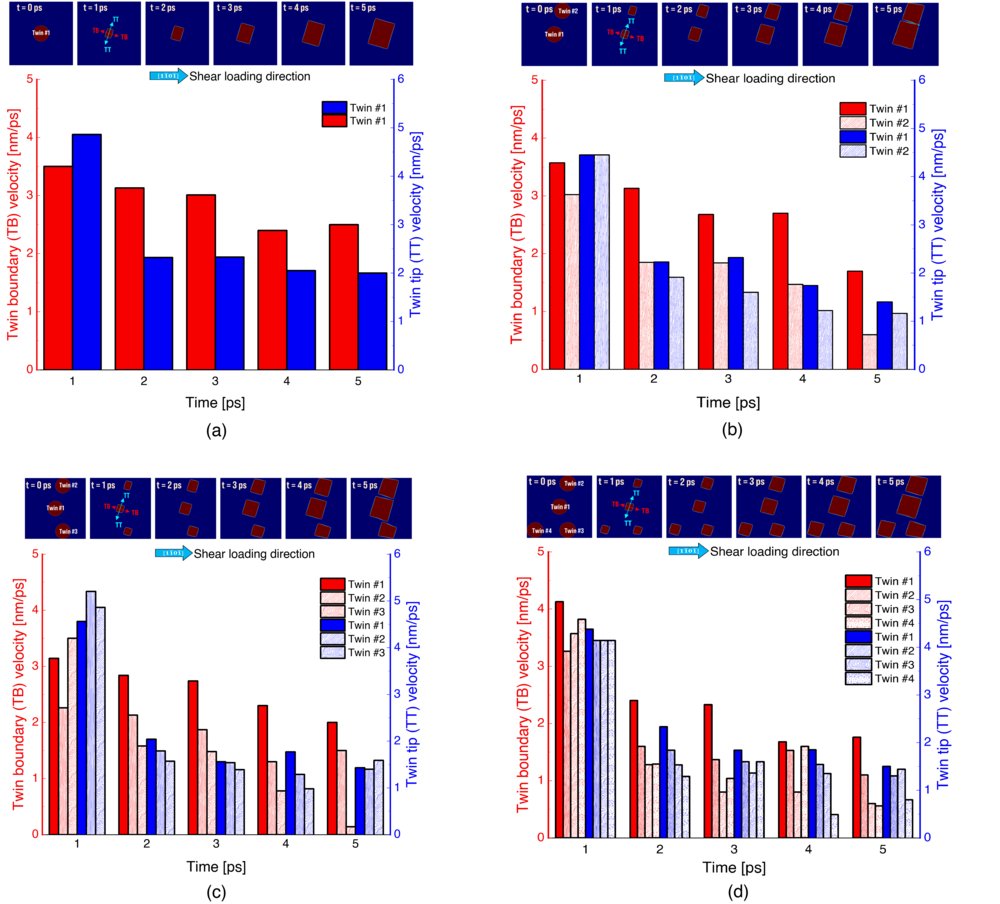}
\caption{Bar charts showing the twin boundary (red arrow) and twin tip (blue arrow) velocities for a single crystal boron carbide by considering different numbers of twin embryos under a shear loading of 0.3: (a) The velocities of a single twin in the center of the numerical geometry at various noted time steps. The insets show the evolution of the twin, parallel and orthogonal to the habit plane; (b) The velocities of two nuclei with respect to time. In the inset, the second twin is inserted at $x = \unit[25]{nm}$ and $y = \unit[35]{nm}$; (c) The velocities are shown for three twin embryos. A different growth direction for the third twin is clear in the inset; (d) The change of twin boundary and twin tip velocities for four nucleus. The growth of each embryo is illustrated in the inset. (For interpretation of the references to color in this figure, the reader is referred to the web version of this article.)}
\label{fig.6}
\end{figure}

Following this basic validation for boron carbide with results under restrictions of experimental limitations in the literature, the change of the twin size (e.g., length and thickness) and twin interactions in a single crystal boron carbide are explored in order to measure the velocity of twin tips and boundaries (Fig.\,\ref{fig.6}). Being an important parameter for indicating the twin boundary propagation as a key plasticity mechanism, the present findings have important implications for studying the morphology of twins. In order to accomplish this endeavor, the velocities are calculated by tracking the mid points ($\eta = 0.5$) on the twin tip and twin boundary interfaces with respect to time. Currently, there is no such statistical data on twin boundary velocity for single crystalline boron carbide, and so we make an attempt to provide some new insights. Considering only one nucleus in the center of the domain, the twin boundary (red colored) and twin tip (blue colored) velocities are shown in Fig.\,\ref{fig.6}(a). The distribution of the twin order parameter at different steps along with the direction for twin tip and twin boundary are also shown in the inset, where the applied shear loading of 0.3 is in the $[1 \overline{1} 0 \overline{1}]$ direction. By choosing $\unit[\Delta t = 1]{fs}$ as the time step, the initial circular nucleus shrinks in size until reaching to a stable shape. After that time, the twin starts to grow in the direction of $73.6^{\circ}$ with respect to the loading direction. In this case, the twin tip and twin boundary velocities are larger at the beginning of the loading in comparison with later time instants due to the detwinning process \cite{doi:10.1063/1.1658166} and larger space for unconfined propagation. In addition, the average of twin tip velocities ($\unit[2.71 \pm 0.86]{\sfrac{nm}{ps}}$) are larger than twin boundaries ($\unit[2.91 \pm 0.37]{\sfrac{nm}{ps}}$) as a result of having a larger aspect ratio. For the two nuclei scenario shown in Fig.\,\ref{fig.6}(b), the average of twin boundary velocities of the middle embryo ($\unit[2.76 \pm 0.48]{\sfrac{nm}{ps}}$) are larger than the single twin case because of the tendency of the middle twin to interact with the twin at the top of the inset (termed as Twin \#2). The variation of the twin tip velocity is also smaller than the single twin case on the basis of the fast growth of the twin's aspect ratio. Moreover, Twin \#2 has a lower aspect ratio, indicating that the two twins will have a wedge shape in the case of interaction between each other. The spreading of a wedge shaped twin has been seen for other ceramics as a result of rapid load drop associated with the twinning process \cite{yangui1982high}.
When placing Twin \#3 at the bottom right of the specimen near the fixed boundary conditions (Fig.\,\ref{fig.6}(c)), the average twin boundary velocities of Twins \#1 and \#2 are increased. This fact is likely a consequence of increasing the twins' aspect ratio, which can be related to the high tendency of twins to interact. Moreover, Twin \#3 grows in the direction perpendicular to other embryos because of arresting at the boundary in the scenario depicted for the three twin systems in Fig.\,\ref{fig.6}(c). By adding another embryo close to the fixed boundary condition in a four twin system (Fig.\,\ref{fig.6}(d)), all the twins' aspect ratio has decreased, with Twin \#2 by $\sim$30\% in both length and width. Furthermore, the embryo in the middle tended to connect to the nucleus at the top of the domain as a result of the proximity of Twin \#2 with the shear loading. Altogether, adding more twin nuclei leads to decreasing the twin boundary velocity of Twin \#1, which may be caused by the local stress created from other nuclei to restrict the movement of the boundary.

\subsection{Fracture-induced twinning in single crystal magnesium}\label{simulation2}

\begin{figure}[H]
\centering
\includegraphics[width=63mm, height=63mm]{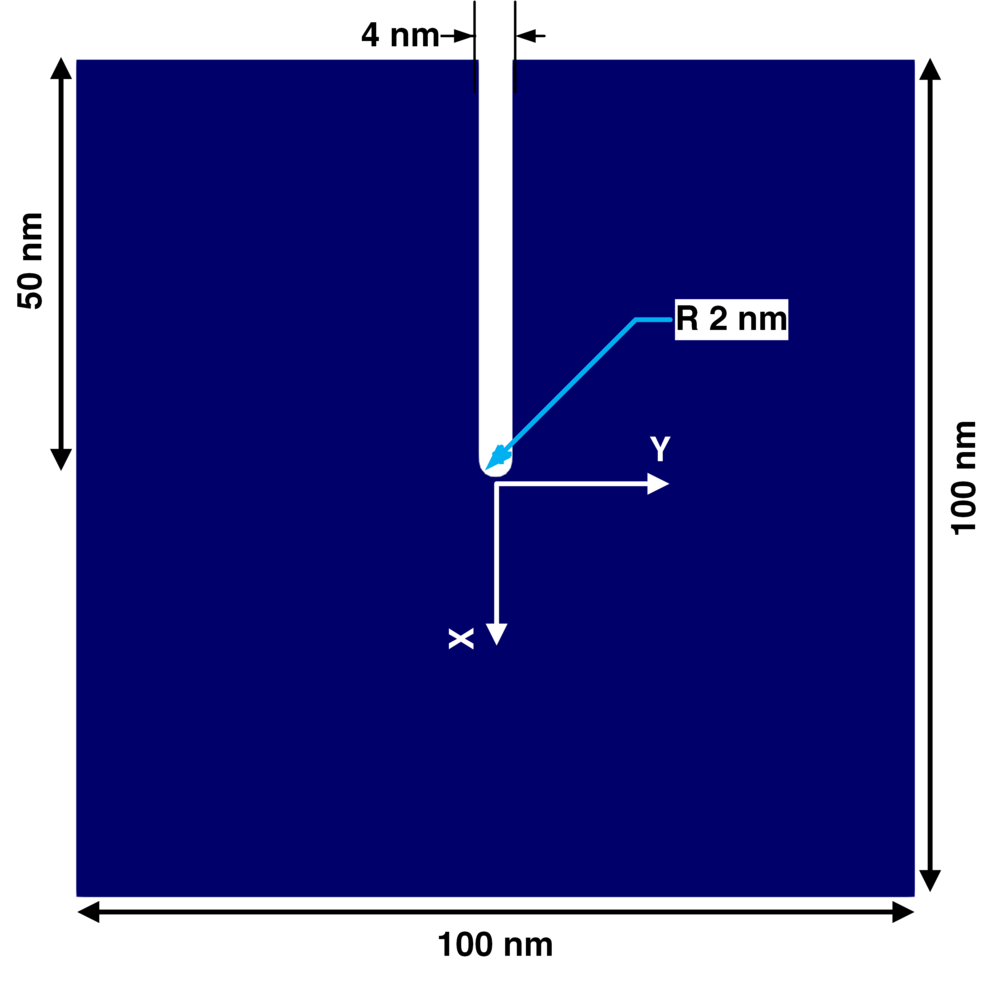}
\caption{A square domain containing an edge crack for numerical simulations under plain strain conditions. The origin of the ($X$, $Y$) coordinate system is at the crack tip, with positive $X$ downward and positive $Y$ to the right.}
\label{fig.7}
\end{figure}

\begin{figure}[H]
\centering
\includegraphics[width=0.8\textwidth, height=115mm]{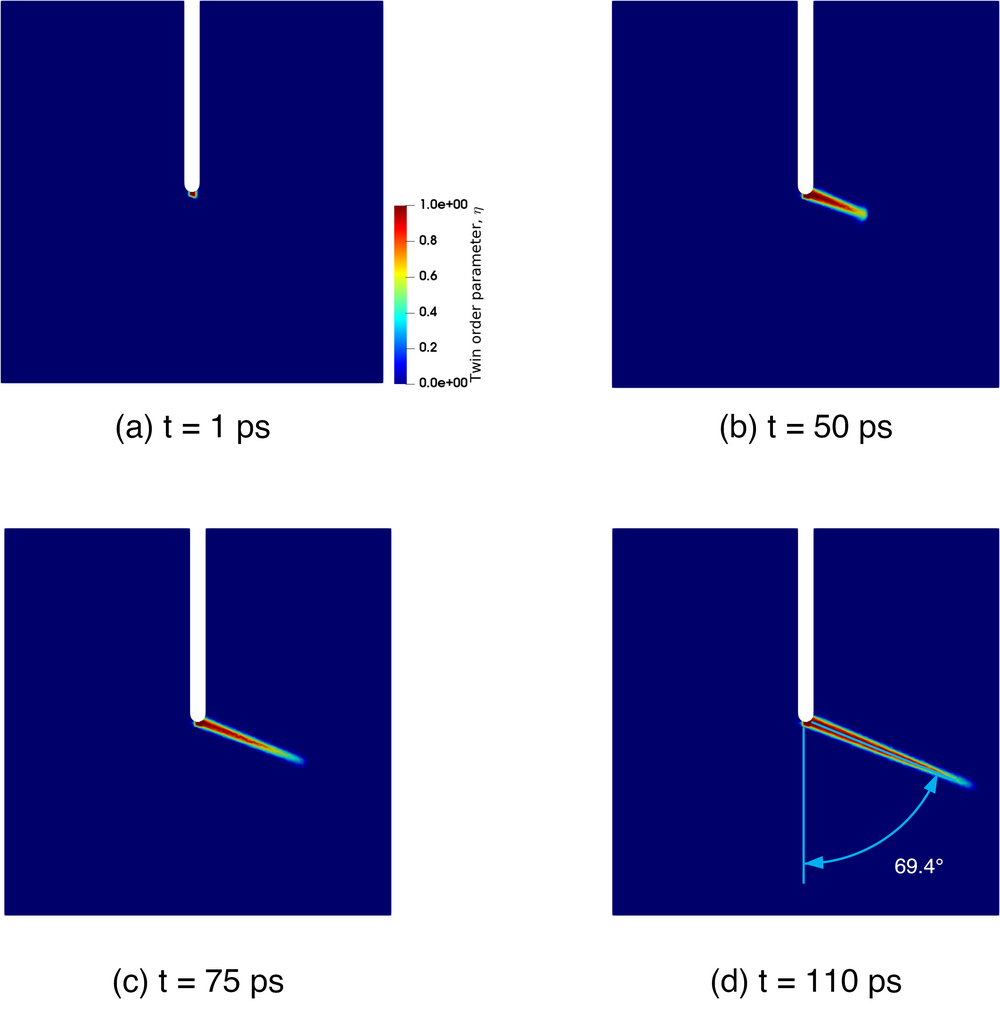}
\caption{Time-evolved twin morphology for mode \Romannum{1} loading of single crystal magnesium at 5\% tensile strain for noted times of: (a) $t = \unit[1]{ps}$, (b) $t = \unit[50]{ps}$, (c) $t = \unit[75]{ps}$, and (d) $t = \unit[110]{ps}$. The resulting twin propagation angle is $69.4^{\circ}$, which is close to the molecular dynamics result of $\sim$ $69^{\circ}$ \cite{tang2011atomistic}. (For interpretation of the references to color in this figure, the reader is referred to the web version of this article.)}
\label{fig.8}
\end{figure}

\begin{figure}[H]
\centering
\includegraphics[width=0.8\textwidth, height=115mm]{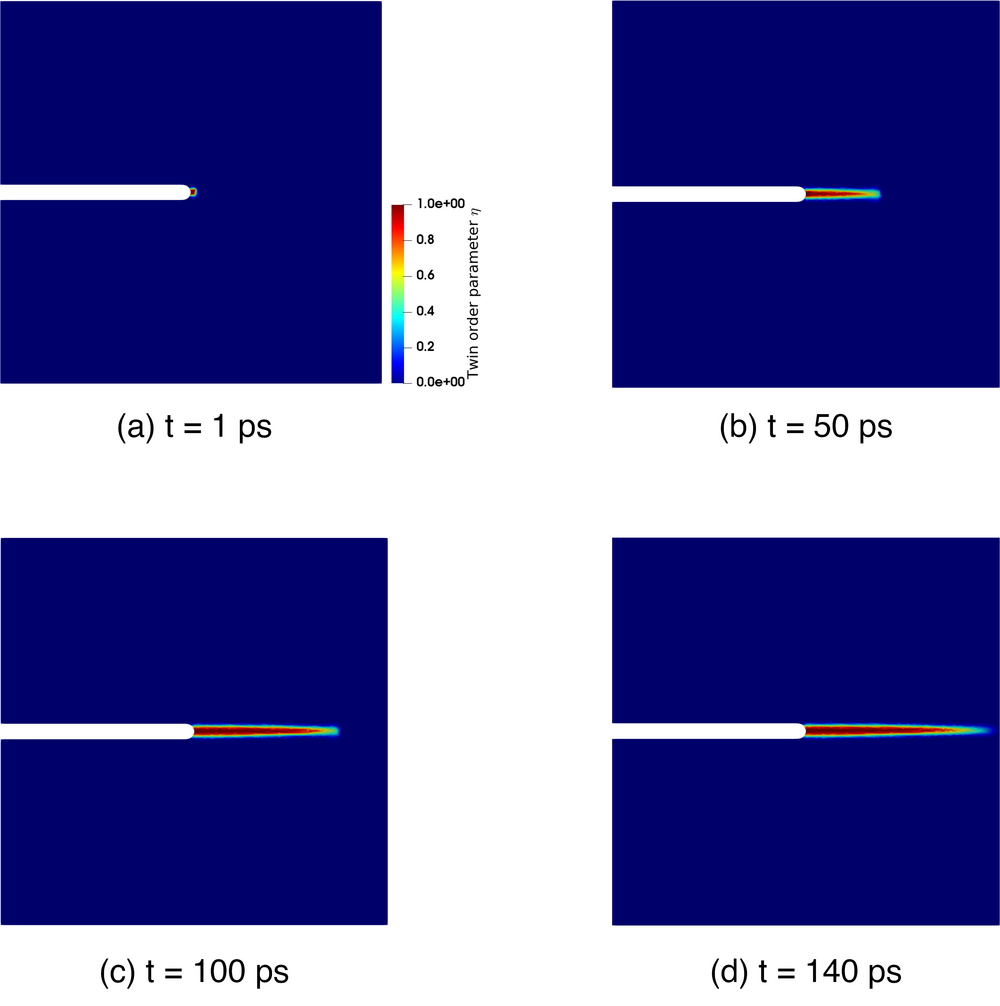}
\caption{Time-evolved order parameter for mode \Romannum{2} loading of a single crystal magnesium at 5\% tensile strain: (a) $t = \unit[1]{ps}$, (b) $t = \unit[50]{ps}$, (c) $t = \unit[100]{ps}$, and (d) $t = \unit[140]{ps}$. (For interpretation of the references to color in this figure, the reader is referred to the web version of this article.)}
\label{fig.9}
\end{figure}

The next example seeks to evaluate the current phase-field approach for studying twinning at a crack tip in magnesium. This simulation is motivated by the urge in understanding the sequence and competition between twinning and fracture, which is difficult to unravel experimentally (e.g., via nanoindentation tests \cite{dehm2018overview}). This is important because far less attention has been given to nucleation and propagation of twins at crack tips and it could offer valuable information on the deformation twinning processes and help to elucidate the role of nanotwinning in crack propagation \cite{yoo1981slip}. In this subsection, a stationary pre-existing crack is considered by a thin notch in a two-dimensional geometry for studying twinning under mode \Romannum{1} and mode \Romannum{2} cracking. The numerical setup is shown in Fig.\,\ref{fig.7}. An initially square domain of size $\unit[100]{nm}$ by $\unit[100]{nm}$ with a pre-existing edge crack of length $\unit[50]{nm}$ and thickness $\unit[4]{nm}$ with a rounded tip of radius $\unit[2]{nm}$ is considered for simulations under a plain strain condition. The crack is assigned a finite radius to alleviate extreme deformations due to singular stress fields at the tip \cite{nakamura1989antisymmetrical}.
For boundary conditions, the crack surface is a free surface with a zero Neumann boundary condition. Along each external boundary condition except for the crack surface, the displacements for pure mode \Romannum{1} or mode \Romannum{2} loading are imposed as in \cite{rice1968mathematical}. The orientation of the twin system, $\boldsymbol{s}$ and  $\boldsymbol{m}$, is chosen such that the resolved shear stress is maximum (i.e., $\theta = \unit[1.2]{rad}$ for mode \Romannum{1} and $\theta = \unit[0]{rad}$ for mode \Romannum{2}). In addition, a small twin nucleus with a radius of $\unit[0.8]{nm}$ at the crack tip is considered as the initial condition for the twin order parameter. The phase-field results for mode \Romannum{1} loading are illustrated in Fig.\,\ref{fig.8}, where a contour of the twin order parameter is plotted. It is clear that the twin growth to the external boundaries is prohibited by the imposed displacement boundary conditions. By progressing in time, the \hkl <1 0 -1 1>\hkl{-1 0 1 2} twin band is nucleated at the crack tip and develops at an externally applied strain of 5\% due to the stress concentration. The shape and angle of the twin of $69.4^{\circ}$ at $t = \unit[110]{ps}$ are in agreement with the atomistic simulation results of tensile twinning in single crystal magnesium \cite{tang2011atomistic}, where a value of $69^{\circ}$ has been reported.
The mode \Romannum{2} case is shown in Fig.\,\ref{fig.9} for the twin order parameter at various time instants. Similar to the mode \Romannum{1} case, the twin nucleates at the crack tip and starts to grow until it is inhibited by the right boundary condition. As expected, the twin system is aligned in a direction that has the maximum resolved shear stress ($\theta = 0$). These results are in qualitative agreement with the stationary phase-field model under similar boundary conditions \cite{clayton2013phase}. This needle-shaped lenticular twin, which has also been observed in \cite{barnett2007twinning}, suggests that twin growth occurs by extension of a fast twin tip followed by a coordinated slower migration of the boundaries \cite{kannan2018mechanics}.

\subsection{Phase-field modeling of fracture subjected to shear and compressive loading in anisotropic boron carbide single crystals}\label{simulation3}

The subject of crack growth in the literature has mainly focused on mode \Romannum{1} fracture because opening mode crack growth is preferred before that under mixed mode or pure shear mode conditions \cite{broberg1987crack}. It is recognized that, even under pure shear loading, local tensile stresses at the tip result in crack growth under mode \Romannum{1} conditions \cite{hayashi1981energy}. However, cracks can grow in brittle materials under mode \Romannum{2} loading when the ratio between the critical stress intensity factors, $\sfrac{K_{\Romannum{2}c}}{K_{\Romannum{1}c}}$, is low \cite{melin1986fracture}. It is also motivated that at a sufficiently high confining pressure, the crack is assumed to extend along a smooth curved path that maximizes $K_{\Romannum{2}}$ \cite{bobet1998fracture}. In heterogeneous brittle solids, the different microstructural inhomogenities (e.g., voids and microcraks) result in a large process regions at the crack tip, and this may lead to macroscopic mode \Romannum{2} failure under compressive loads \cite{jung1995study}. The study of crack initiation and propagation of mode \Romannum{2} fracture is, thus, important in order to better understand the behavior of cracks in brittle solids.

Classically in phase-field modeling in the literature \cite{bleyer2018phase}, it is assumed that for compressive deformation states, crack growth does not take place. To deal with this, a common technique is to decompose the strain energy density into tensile and compressive parts using a spectral decomposition \cite{miehe2010thermodynamically}, or a hydrostatic-deviatoric approach \cite{amor2009regularized}; however, both of these decompositions have disadvantages that have yet to be addressed. Specifically, regarding the spectral decomposition, the force-displacement curve shows unphysical stiffening in the fully-cracked specimen \cite{cajuhi2018phase}. For the hydrostatic-deviatoric method, there are limitations for compression-dominated loading (e.g., the material is allowed to crack in volumetric expansion and shear, but not in volumetric compression) \cite{sargado2018high}. In addition, both of these popular decompositions can only be used for isotropic materials \cite{van2020strain}. Nevertheless, boron carbide has strong anisotropic elasticity ($\frac{E_{\text{max}}}{E_{\text{min}}} = 8.11$, where $E_{\text{max}}$ and $E_{\text{min}}$ are the general maximum and minimum Young's modulus, respectively) \cite{mcclellan2001room}. This analysis is important because the plastic deformation in nanograined boron carbide is assumed to be dominated by intergranular fracture \cite{https://doi.org/10.1111/jace.18071} and these new results can be employed toward guiding material design for \ch{B4C} under extreme dynamic loading.

\subsubsection{Crack initiation and propagation under biaxial compressive stress in single crystal boron carbide}

\begin{figure}[H]
\centering
\includegraphics[width=85mm, height=85mm]{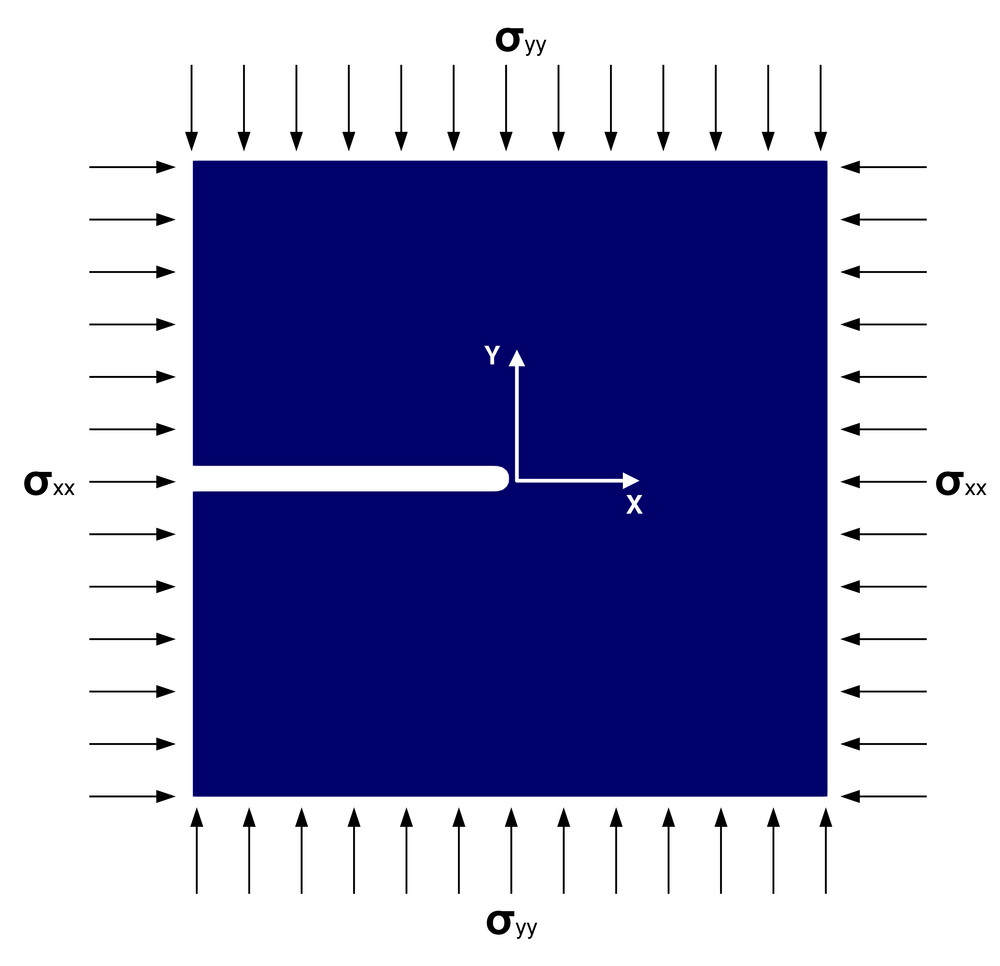}
\caption{Geometry and boundary conditions in the numerical simulations of biaxial loading of boron carbide in a pre-notched domain. The Cartesian coordinate system is considered at the crack tip.}
\label{fig.10}
\end{figure}

\begin{figure}[H]
\centering
\includegraphics[width=0.8\textwidth, height=110mm]{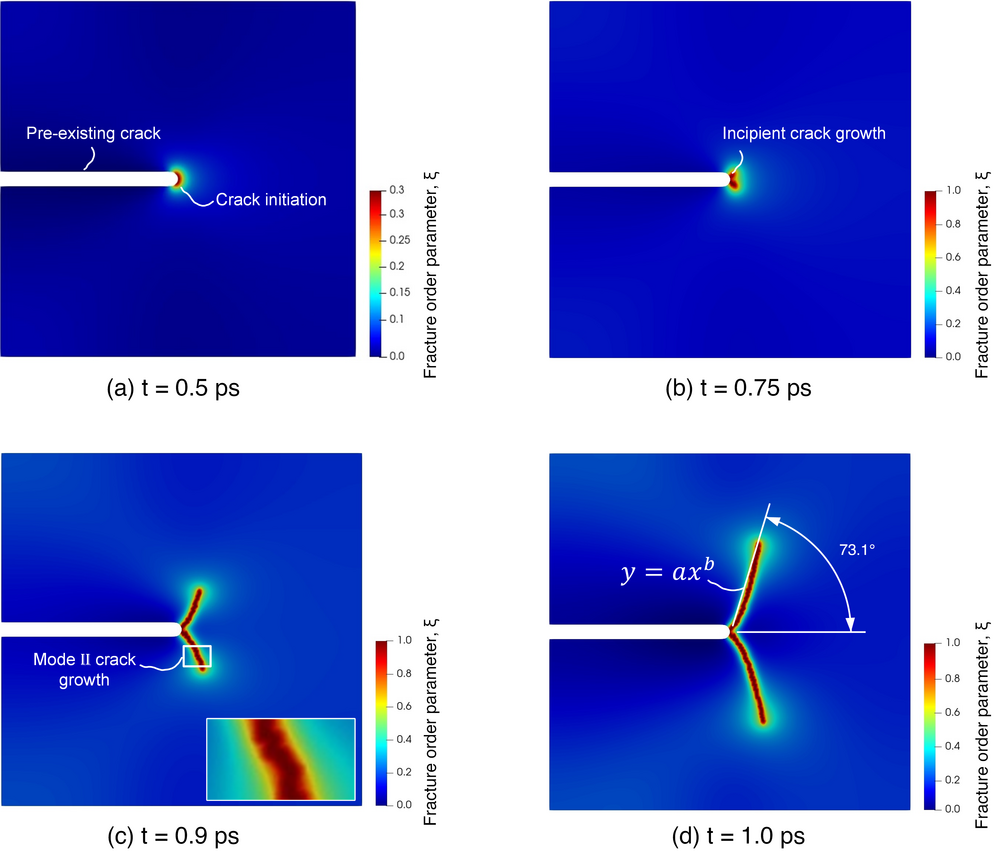}
\caption{Crack evolution obtained with the phase-field approach under a biaxial stress loading condition in single crystal \ch{B4C}: (a) At $t = \unit[0.5]{ps}$, the fracture order parameter starts to accumulate at the tip of the pre-existing notch. The maximum value of the order parameter is 0.3, showing that the crack region is not fully formed; (b) Two fully cracked regions start to grow via an incipient kink at $t = \unit[0.5]{ps}$; (c) Two anti-symmetric cracks at $t = \unit[0.9]{ps}$ which emerged first at the crack tip. The inset shows the mode \Romannum{2} crack growth under a combined load of shear and high compression. (d) Stable propagation of cracks along a curvilinear path described by $y = ax^{b}$ at $t = \unit[1.0]{ps}$. The angle from the previous crack plane to the new assumed direction of crack growth is $73.1^{\circ}$. (For interpretation of the references to color in this figure, the reader is referred to the web version of this article.)}
\label{fig.11}
\end{figure}

Consider the biaxial compression test of a single crystal \ch{B4C} specimen with a single pre-existing notch under plain strain condition as shown in Fig.\,\ref{fig.10}. The dimensions of the square domain are those of Fig.\,\ref{fig.7}, and the material parameters are the same as those mentioned in Table \ref{accuracy}. Additionally, the fracture surface energy ($\Upsilon$) and cleavage anisotropy factor ($\beta$) are set to $\unit[3.27]{}$ and $\unit[100]{\sfrac{J}{m^2}} \ (\text{or} \ 0 \ \text{for isotropic damage})$, respectively \cite{beaudet2015surface}. In the simulations, a total of 323,460 triangular elements were used to discretize the domain with a finer mesh assigned to critical zones. A high confining stress is chosen such that the opening stress intensity factors at the tip of the crack in any direction is zero. The stress parallel to the crack plane is assumed to be larger than the stress values normal to the crack plane ($\sigma_{XX} > \sigma_{YY}$). The initial condition for the time-dependent fracture order parameter and time step are set to $\xi(t=0) = \unit[0.01]{}$and $\Delta t = \unit[0.5]{fs}$, respectively. In the simulations, all the frictional effects on the crack surfaces are disregarded.
The crack evolution process under these numerical conditions is depicted in Fig.\,\ref{fig.11}. As shown, biaxial compression first leads to the initiation of the crack from the tip of the notch (Fig.\,\ref{fig.11}(a)). The range of the fracture order parameter indicates that the crack is not fully formed at $t = \unit[0.5]{ps}$. At $t = \unit[0.75]{ps}$ (Fig.\,\ref{fig.11}(b)), the crack kinks as two single straight branched cracks at a small angle. By progressing in time to $t = \unit[0.9]{ps}$, two anti-symmetric cracks begin to propagate toward the top and bottom boundaries due to the larger compressive normal stress parallel with the crack plane (Fig.\,\ref{fig.11}(c)). In addition, the crack grows in incrementally small steps that are consistent with experimental observations for other brittle materials \cite{leblond2000crack, bobet2000initiation}. At the last time frame of $t = \unit[1]{ps}$, the propagation path of cracks in single crystal \ch{B4C} is shown (Fig.\,\ref{fig.11}(d)). As can be seen, the crack patterns follow a curvilinear path described by a function $ax^{b}$. The crack paths reported analytically in \cite{isaksson2002mode} and measured experimentally in \cite{leblond2000crack} support this computational result herein. In the studied experiments, $b$ has been found in the interval of 1.43 to 1.58 for pre-fractured specimens of gypsum under uniaxial and biaxial compression \cite{bobet1998fracture}. From the analytical model, the exponent was required to be equal to 1.5 in order to be independent of the crack extension length \cite{isaksson2002mode}. For boron carbide in this study, the exponent $b$ is obtained as $1.66 \pm 0.06$ and this is in reasonable agreement with the predicted theory for brittle materials. The curvature parameter $a$ is equal to $0.49 \pm \unit[0.08]{\left(nm\right)^{-0.66}}$ and the angle of the branched kink is $73.1^{\circ}$, which is in good agreement with the value ($70^{\circ}$) reported in \cite{melin1986does}. 

Finally, the homogeneous damage distribution ($\xi \sim 0.5$) in Fig.\,\ref{fig.11} is also due to the hydrostatic nature of the loading. As it is shown in \cite{levitas2018thermodynamically}, the damage initiation criteria in the current phase-field potential for fracture is fulfilled at infinitesimal load; however, at the crack free surface where the load is not applied, the color is dark blue, which indicates no damage, as expected.

\section{Conclusion} \label{section.conclusion}

A robust finite element procedure for solving a coupled system of equilibrium and time-dependent Ginzburg--Landau equations has been motivated by using thermodynamically-sound derivation of governing equations. Use of the variational procedure and thermodynamically consistency of the model ensures that it has a strict relaxational behaviour of the free energy; hence, the models are more than a phenomenological description of an interfacial problem as was done previously in the literature \cite{guo2015thermodynamically}. The dissipation and time scales associated with growth kinetics are also derived and addressed in our paper. The model has been used for studying the evolution of twinning deformation and fracture in anisotropic single crystal magnesium and boron carbide at finite strains. The formulation considers distinct order parameters for fracture and twinning. For the first time, a monolithic strategy has been employed for solving the coupled mechanical equilibrium and order parameters evolution equations under extreme conditions. As a challenge in continuum mechanics, nanometer length scale and picosecond time scale have been used in simulations in this paper.

The computational procedures and numerical algorithms are implemented using the open-source platform FEniCS. The present nonlinear finite element code has been developed and used to study: (i) the growth and propagation of deformation twinning in single crystal magnesium and boron carbide, (ii) fracture-induced twinning in single crystal magnesium under pure mode \Romannum{1} and mode \Romannum{2} loading, and (iii)  the prediction of the crack path under biaxial compressive stress loading in single crystal boron carbide. The numerical results for all the problems are in agreement with the available experimental data and analytical solutions in the literature. It has been demonstrated through numerical simulations that the proposed model delivers adequate results matching qualitatively a variety of observed phenomena, including the growth of existing twin embryos, the effect of pre-existing cracks on the twin path under various loading, and the propagation of cracks under compression for highly anisotropic boron carbide. The current contribution opens up new possibilities for multi-scale fracture models. In the future, our finite element based phase-field model can be applied for studies of phase transformations (e.g., amorphization \cite{clayton2014phase}) and interaction between plasticity and fracture under high strain-rate loading. As a next step, the current model could be combined with discrete localized plastic flow (e.g., shear band and dislocation pileups \cite{javanbakht2015interaction}) and thermally-activated mechanisms (e.g., melting \cite{levitas2015multiphase}) to capture the behavior of brittle materials in laser spall experiments.

\section{Data Availability}

The authors declare that the main data supporting the findings of this study are available within this article. Extra data are available from the corresponding authors upon reasonable requests.

\section{Code Availability}

The Python code, generated during the current study, is part of the FEniCS project available at \href{http://www.fenicsproject.org/download}{http://www.fenicsproject.org/download}, and an example for the computational implementation is available in \cite{compreal} to be used under the GNU Public license \cite{gnupublic}.

\section{Declaration of Competing Interests}

The authors declare no competing financial interests or personal relationships.

\section{CRediT Authorship Contributions Statement}

B.A. developed the model, wrote the code, designed and performed all simulations, analyzed results, and wrote the original draft. H.J. analyzed results, reviewed, and edited the paper. B.E.A. developed the model, helped with the code, allocated the computational resources, reviewed and edited the paper. A.R. helped in computational aspects, reviewed and edited the paper. J.D.H. supervised the research, acquired funding, reviewed, and edited the paper. All authors discussed the results.

\section{Acknowledgements}

B.A. and J.D.H acknowledge support from Natural Sciences and Engineering Research Council of Canada (NSERC) Discovery Grant 2016-04685 and NSERC DNDPJ 531130-18. H.J. and A.R. acknowledge the partial support of the MIUR-PRIN project XFAST-SIMS (no. 20173C478N). HJ was funded by the Alexander von Humboldt Foundation during his stay at ICAMS.

\bibliographystyle{special}
\bibliography{Benhour} 

\end{document}